\documentclass{ar2e}

\usepackage{graphicx}
\usepackage{amsmath}
\usepackage{bm}
\usepackage{cite}

\newcommand{\bqa}{\begin{eqnarray}}
\newcommand{\eqa}{\end{eqnarray}}
\newcommand{\be}{\begin{equation}}
\newcommand{\ee}{\end{equation}}
\newcommand{\ba}{\begin{eqnarray}}
\newcommand{\ea}{\end{eqnarray}}
\newcommand{\beq}{\begin{equation}}
\newcommand{\eeq}{\end{equation}}
\newcommand{\beqa}{\begin{eqnarray}}
\newcommand{\eeqa}{\end{eqnarray}}

\newcommand{\sing}{^1{\rm S}_0}
\newcommand{\trip}{^3{\rm S}_1}

\newcommand{\mpic}{M_\pi^{crit}}
\newcommand{\Mhi}{M_{high}}
\newcommand{\Mlo}{M_{low}}

\begin{document}



\jname{Annual Review in Nuclear and Particle Science}
\jyear{}
\jvol{}
\ARinfo{}

\title{
Efimov States in Nuclear and Particle Physics}

\markboth{Hammer, Platter}{Efimov States in Nuclear and Particle Physics}

\author{Hans-Werner Hammer\footnote{Corresponding author}
\affiliation{Helmholtz-Institut f\"ur Strahlen- und Kernphysik
        (Theorie) and Bethe Center for Theoretical Physics,
        Universit\"at Bonn, 53115 Bonn, Germany\\
        Email: hammer@hiskp.uni-bonn.de}
Lucas Platter
\affiliation{Institute for Nuclear Theory, University of Washington, Seattle,
WA\ 98195, USA\\
Department of Physics,
         The Ohio State University, Columbus, OH\ 43210, USA\\
Email: lplatter@phys.washington.edu}
}

\begin{keywords}
Universality, Efimov physics, few-body systems, 
discrete scale invariance, hyperspherical formalism, effective field theory
\end{keywords}

\begin{abstract}
\baselineskip 12pt 
Particles with resonant short-range interactions 
have universal properties 
that do not depend on the details of their structure or their
interactions at short distances. In the three-body system, these properties 
include the existence of a geometric spectrum of three-body Efimov states
and a discrete scaling symmetry leading to 
log-periodic dependence of observables on the scattering length.
Similar universal properties appear in the four-body system and 
possibly higher-body systems as well. For example, universal four-body
states have recently been predicted and observed in experiment.
These phenomena are often referred to as \lq\lq Efimov Physics''.
We review their theoretical description and discuss 
applications in different areas of physics with a special emphasis on 
nuclear and particle physics.
\end{abstract}

\maketitle
\topmargin 2cm

\baselineskip 18pt 
\newpage
\section{Introduction}
\label{sec:introduction}
The scattering of particles with sufficiently low kinetic energy is
determined by their S-wave scattering length $a$.  This is the case if
their de Broglie wavelengths are large compared to the range of the
interaction.

Generically, the scattering length $a$ is comparable in magnitude to
the range $\ell$ of the interaction: $|a| \sim \ell$.  In exceptional
cases, the scattering length can be much larger in magnitude than the
range: $|a| \gg \ell$.  Such a large scattering length requires the
fine-tuning of a parameter characterizing the interactions to the
neighborhood of a critical value at which $a$ diverges to $\pm
\infty$.  If the scattering length is large, the particles exhibit
properties that depend on $a$ but are insensitive to the range and
other details of the short-range interaction.  These properties are
universal in the sense that they apply equally well to any
nonrelativistic particle with short range interactions that produce a
large scattering length \cite{Braaten:2004rn,Platter:2009gz}.

For example, in the case of equal-mass particles with 
mass $m$ and $a>0$, there is
a two-body bound state near the scattering threshold with binding energy
$B_d = \hbar^2 /(m a^2)$.  The corrections to this formula are
suppressed by powers of $\ell /a$.  This bound state corresponds to a
pole of the two-particle scattering amplitude at $E=-B_d$.  If the
scattering length is negative, there is a universal virtual state
which corresponds to a pole on the unphysical second sheet in the
complex energy plane.

The key evidence for universal behavior in the three-body system was the
discovery of the Efimov effect in 1970 \cite{Efimov-70}.  In the
unitary limit $1/a \to 0$, the two-body bound state is exactly at the
two-body scattering threshold $E = 0$.  Efimov showed that in this limit
there are infinitely many, arbitrarily-shallow three-body bound states
whose binding energies $B_t^{(n)}$ have an accumulation point at $E =
0$.  The Efimov effect is just one aspect of universal properties in
the three-body system.  It has universal properties not only in the
unitary limit, but whenever the scattering length is large compared to
the range $\ell$. In particular, the log-periodic scattering length 
dependence of observables is a unique consequence of Efimov physics.

Although well established theoretically,
the umambiguous identification of Efimov states in nature is 
difficult since typical systems are neither in the unitary limit nor
can the scattering length be varied.
The probably simplest example in nuclear physics is the triton.
The triton can be interpreted as the ground state of an Efimov 
spectrum in the $pnn$-system with total spin $J=1/2$. Since the 
ratio $\ell/a$ is only about one third, the whole spectrum contains only
one state but the low-energy properties of the triton can be described
in this scenario.
A promising system for observing several Efimov states is $^4$He
atoms, which have a scattering length that is more than a factor of 10
larger than the range of the interaction.  Calculations using accurate
potential models indicate that the system of three $^4$He atoms has
two three-body bound states or trimers.  The ground-state trimer can be
interpreted as an Efimov state, and it has been observed in
experiments involving the scattering of cold jets of $^4$He atoms from
a diffraction grating \cite{STo96}.  The excited trimer is universally
believed to be an Efimov state, but it has not yet been observed.

The rapid development of the field of cold atom physics has opened up
new opportunities for the experimental study of Efimov physics.  This
is made possible by two separate technological developments.  One is
the technology for cooling atoms to the extremely low temperatures
where Efimov physics plays a crucial role.  The other is the
technology for controlling the interactions between atoms.  By tuning
the magnetic field to a Feshbach resonance, the scattering lengths of
the atoms can be controlled experimentally and made arbitrarily large.
Both developments were crucial in recent experiments that provided the
first indirect evidence for the existence of Efimov states in
ultracold atoms \cite{Kraemer:2006}.

Overviews of Efimov physics in ultracold atomic gases can be found in
Refs.~\cite{Braaten:2004rn,Braaten:2006vd,Platter:2009gz}.  
In this review, we focus
on universal aspects and Efimov states in nuclear and particle
physics.  Even though the scattering length can not be varied, there
are many systems close to the unitary limit where Efimov physics is
relevant. They include few-nucleon systems, halo nuclei, and weakly
bound hadronic molecules.  These systems can be described in a
universal effective field theory (EFT) that implements an expansion
around the unitary limit. Three-body bound states can be interpreted
as Efimov trimers.

In the next section we will review the physics of the Efimov effect
starting with a brief account of the history.  In the following
sections, we will discuss applications in nuclear and particle
physics. We will end with a summary and outlook.

\section{Physics of the Efimov Effect}
\label{sec:phys-efim-effect}
\subsection{History}
\label{sec:history}
The first hints on universal behavior in the three-body system came
from the discovery of the Thomas collapse in 1935 \cite{Tho35} which is
closely related to the Efimov effect.  Thomas studied the zero-range
limit for potentials with a single two-body bound state with fixed
energy.  Using a variational argument, he showed that the binding
energy $B_t^{(0)}$ of the deepest three-body bound state diverges to
infinity in this limit.  Thus the spectrum of three-body bound states
is unbounded from below.

Further progress occured by applying the zero-range limit to the
three-nucleon system.  An integral equation for S-wave
neutron-deuteron scattering using zero-range
interactions was derived by Skorniakov and Ter-Martirosian in 1957
\cite{Skorniakov:1957aa}. For the spin-quartet channel, this integral
equation has no bound state solutions and is well behaved.  In the
spin-doublet channel, however, it has solutions for arbitrary energy
\cite{Danilov:1961aa}, including bound state solutions. If the
solution is fixed by requiring some three-body energy, the resulting
equation still has a discrete spectrum that extends to minus infinity
in agreement with the earlier result of Thomas
\cite{Faddeev:1961bg,MiF62}.  Although a prediction for the
spin-doublet neutron-deuteron scattering length was obtained using the
triton binding energy as input \cite{DaL63}, most work afterwards
focused on finite range forces which avoid this pathology at high
energies.

In 1970, Efimov realized that one should focus on the physics at low
energies, $E \ll \hbar^2/(m\ell^2)$, and not on the deepest states.  
In this limit,
where zero-range forces are adequate, he found some surprising results
\cite{Efimov-70}.  He pointed out that when $|a|$ is sufficiently
large compared to the range $\ell$ of the potential, there is a
sequence of three-body bound states whose binding energies are spaced
roughly geometrically in the interval between $\hbar^2/(m \ell^2)$ and
$\hbar^2/(m a^2)$.  As $|a|$ is increased, new bound states appear in
the spectrum at critical values of $a$ that differ by multiplicative
factors of $e^{\pi/s_0}$, where $s_0$ depends on the statistics and
the mass ratios of the particles.  In the case of spin-doublet
neutron-deuteron scattering and for three identical bosons, $s_0$ is
the solution to the transcendental equation
\begin{eqnarray}
s_0 \cosh {\pi s_0 \over 2} = {8 \over \sqrt{3}} \sinh {\pi s_0 \over 6} \,.
\label{s0}
\end{eqnarray}
Its numerical value is $s_0\approx 1.00624$, so $e^{\pi/s_0} \approx
22.7$.  As $|a|/\ell \to \infty$, the asymptotic number of three-body bound
states is
\begin{eqnarray}
N \longrightarrow {s_0 \over \pi} \ln {|a| \over \ell} \,.
\label{N-Efimov}
\end{eqnarray}
In the limit $a \to \pm \infty$, there are infinitely 
many three-body bound states with an accumulation point at the 
three-body scattering threshold with a geometric spectrum:
\begin{eqnarray}
B^{(n)}_t = (e^{-2\pi/s_0})^{n-n_*} \hbar^2 \kappa^2_* /m,
\label{kappa-star}
\end{eqnarray}
where $m$ is the mass of the particles and
$\kappa_*$ is the binding wavenumber of the branch of Efimov states
labeled by $n_*$.  The geometric  spectrum in (\ref{kappa-star})
is the signature of a
discrete scaling symmetry with scaling factor $e^{\pi/s_0}\approx 22.7$.
It is independent of the mass or structure of the identical
particles and independent of the form of their short-range interactions. 
The Efimov effect can also occur in other three-body systems if at least two 
of the three pairs have a large S-wave scattering length
but the numerical value of the asymptotic ratio
may differ from the value 22.7.  

A formal proof of the Efimov effect was subsequently given by Amado
and Noble \cite{Amado197125,PhysRevD.5.1992}.  The Thomas and Efimov
effects are closely related. The deepest three-body bound states found by
Thomas's variational calculation can be identified with the deepest
Efimov states \cite{PhysRevA.37.3666}.
The mathematical connection of the Efimov effect to a limit cycle was 
discussed in \cite{Albeverio1981105}.

The universal properties in the three-body system with large scattering
length are not restricted to the Efimov effect.  The dependence of
three-body observables on the scattering length or the energy is
characterized by scaling behavior modulo coefficients that are
log-periodic functions of $a$ \cite{Efimov:1971zz,Efimov:1978pk}.
This behavior is characteristic of a system with a discrete scaling
symmetry.  We will refer to universal aspects associated with a
discrete scaling symmetry as Efimov physics.

In 1981, Efimov proposed a new approach to the low-energy few-nucleon
problem in nuclear physics that, in modern language, was based on
perturbation theory around the unitary limit \cite{Efimov:1981aa}.
Remarkably, this program works reasonably well in the three-nucleon system
at momenta small compared to $M_\pi$.
The Efimov effect makes it necessary to impose a boundary condition on
the wave function at short distances. The
boundary condition can be fixed by using either the spin-doublet
neutron-deuteron scattering length or the triton binding energy as
input. If the deuteron binding energy and the spin-singlet 
nucleon-nucleon scattering
length are used as the two-body input and
if the boundary condition is fixed by using the spin-doublet
neutron-deuteron scattering length as input, the triton binding energy
is predicted with an accuracy of 6\%. The accuracy of the predictions
can be further improved by taking into account the effective range as
a first-order perturbation \cite{Efimov:1991aa}. Thus the triton can
be identified as an Efimov state associated with the deuteron 
and the spin-singlet virtual state being a
$pn$ state with large scattering length \cite{Efimov:1981aa}.

In the three-nucleon system, this program was implemented within an
effective field theory framework by Bedaque, Hammer, and van Kolck
\cite{Bedaque:1997qi,Bedaque:1998mb,Bedaque:1999ve}. In
Ref.~\cite{Bedaque:1999ve}, they found that the renormalization of the
effective field theory requires a $SU(4)$-symmetric three-body interaction
with an ultraviolet limit cycle. The three-body force depends on a
parameter $\Lambda_*$ that is determined through a renormalization
condition that plays the same role as Efimov's boundary condition.
$SU(4)$-symmetry was introduced by Wigner in 1937 as generalization of
the $SU(2)\times SU(2)$ spin-isospin symmetry, allowing for a mixing
of spin and isopin degrees of freedom in symmetry transformations
\cite{Wigner37}. It is satisfied to a high degree in the energy
spectra of atomic nuclei. Exact Wigner symmetry requires the S-wave
scattering lengths in the spin-triplet and spin-singlet 
channels to be equal. However, if the
two-body scattering lengths are large, it is a very good approximation
even if they are different since the symmetry-breaking terms are
proportional to the inverse scattering lengths \cite{Mehen:1999qs}.

This effective field theory (EFT) is ideally suited to calculating 
corrections to the universal results in the scaling limit. 
Its application to few-nucleon physics will be discussed 
in Section \ref{sec:nucphys}.

\subsection{Hyperspherical Methods} 
\label{sec:hypersph-meth}
Coordinate space methods have proven to be a valuable tool for the
analysis of the three-body problem with short-range interactions
\cite{Efimov:1971zz,NFJG01,Jensen-04}. In
this subsection we will introduce the hyperspherical approach that has
been used to obtain important results about the Efimov spectrum. The
material of this subsection is based on the discussion of the
hyperradial formalism in the review of Ref.~\cite{Braaten:2004rn}. For
three particles of equal mass the Jacobi coordinates are defined as:
\begin{equation}
{\bf r}_{ij}={\bf r}_i - {\bf r}_j~;\quad{\bf
    r}_{k,ij}={\bf r}_k - \frac{1}{2}({\bf r}_i + {\bf r}_j)~,
  \label{eq:Jacobi}
\end{equation}
where the triple $(ijk)$ is a
cyclic permutation of the particle indices $(123)$. The hyperradius
$R$ and hyperangle $\alpha_k$ are then defined by \begin{equation}
  R^2=\frac{1}{3}({\bf r}_{12}^2 + {\bf r}_{23}^2 + {\bf
  r}_{31}^2)=\frac{1}{2} {\bf r}_{ij}^2 + \frac{2}{3}{\bf r}_{k,ij}^2; \quad
\alpha_k=\arctan\left(\frac{\sqrt{3} |{\bf r}_{ij}|}{2 |{\bf
    r}_{k,ij}|}\right).
\end{equation}
In the center-of-mass system the Schr\"odinger equation in hyperspherical
coordinates is given by
\begin{equation}
\left(T_R + T_{\alpha_k} + \frac{\Lambda_{k,ij}^2}{2 m R^2} +
V(R,\Omega)\right)
\Psi(R,\alpha,\Omega)=E \Psi(R,\alpha,\Omega),
\label{eq:hyperschro}
\end{equation}
with
\begin{eqnarray}
T_R&=&\frac{\hbar^2}{2 m} R^{-5/2} \left(-\frac{\partial}{\partial R^2} +
\frac{15}{4 R^2}\right) R^{5/2},\\
T_\alpha&=&\frac{\hbar^2}{2 m R^2} \frac{1}{\sin 2 \alpha}
\left(-\frac{\partial^2}{\partial \alpha^2} - 4\right) \sin 2 \alpha,\\
{\Lambda_{k,ij}}^2&=&\frac{\mathbf L_{ij}^2}{\sin^2\alpha_k}  +
\frac{\mathbf L_{k,ij}^2}{\cos^2 \alpha_k},
\label{eq:angmom}
\end{eqnarray}
where $\Omega=(\theta_{ij},\phi_{ij},\theta_{k,ij},\phi_{k,ij})$ and
the $L$s that appear in Eq.~(\ref{eq:angmom}) are the usual
angular-momentum operators with respect to these angles.

We will assume that the potential $V$ depends only on the magnitude of the
inter-particle separation and write
\begin{equation}
V({\bf r}_1,{\bf r}_2,{\bf r}_3)=V(r_{12}) + V(r_{23}) + V(r_{31}).
\label{eq:Vid}
\end{equation}
We now employ the usual Faddeev decomposition of $\psi$ for three
identical bosons and neglect subsystem angular momentum
\begin{equation}
\Psi(R,\alpha,\Omega)=\psi(R,\alpha_1) + \psi(R,\alpha_2) +
\psi(R,\alpha_3).
\label{eq:decomp}
\end{equation}

The solution of the corresponding Faddeev equation can then be expanded 
in a set of eigenfunctions of the hyperangular operator, i.e.
\begin{equation}
  \psi(R,\alpha)=\frac{1}{R^{5/2} \sin(2 \alpha)} \sum_n f_n(R)
  \phi_n(R,\alpha)~.
\end{equation}
This leads to separate differential equations for the hyperangular
functions $\phi_n$ and the hyperradial functions
$f_n$. We obtain in particular for the hyperradial functions
\begin{eqnarray}
\nonumber
Ef_n(R)&=&\left[\frac{\hbar^2}{2m}\left(-\frac{\partial^2}{\partial R^2} + \frac{15}{4
    R^2}\right) + V_n(R)\right]f_n(R)\\
&&
\hspace{20mm}+ \sum_m \left[2 P_{nm}(R) \frac{\partial}{\partial R} +
  Q_{nm}(R)\right] f_m(R),
\label{eq:hyperrad}
\end{eqnarray}
with the hyperradial potential $V_n(R)$ defined by
\begin{equation}
V_n(R)=\left(\lambda_n(R)-4\right)\frac{\hbar^2}{2 m R^2},
\end{equation}
and $P_{nm}(R)$ and $Q_{nm}(R)$ potentials that induce coupling between
different hyperradial channels~\cite{Braaten:2004rn}.

For hyperradii $R$ which are much larger than the range $\ell$ over
which $V$ is non-zero, the solution of the equation for the
hyperangular function $\phi_n$ for large $\alpha$ is
\begin{equation}
  \phi_n^{\rm{(high)}}(\alpha)\approx\sin\left[\sqrt{\lambda_n}\left(\frac{\pi}{2}
    - \alpha\right)\right].
\label{eq:phihigh}
\end{equation}

On the other hand,
for $R \gg \ell$ and small $\alpha$, it can be shown that the solution
for the hyperangular part can be written as
\begin{equation}
\phi_n^{\rm{(low})}(\alpha)=A\,\psi_{0}(\sqrt{2} R \alpha)
-\frac{8
  \alpha}{\sqrt{3}} \sin\left(\sqrt{\lambda_n} \frac{\pi}{6} \right)~,
\label{eq:philow}
\end{equation}
where $A$ is a constant and $\psi_0$ is the zero-energy solution to a
two-body Schroedinger equation with the two-body potential $V$ 
\begin{equation}
  \psi_k(r)=\frac{\sin(kr + \delta(k))}{k}=\frac{\sin
  \delta(k)}{k}[\cos(kr) + \cot \delta \sin(kr)].
\label{eq:psik}
\end{equation}
As $k \rightarrow 0$ this yields $\psi_{0}(r)=r-a$, and we can use this
asymptotic two-body wave function in Eq.~(\ref{eq:philow}). This gives
\begin{equation}
  \phi_n^{\rm (low)}(\alpha)=A (\sqrt{2} R \alpha - a) - \frac{8
  \alpha}{\sqrt{3}}\sin\left(\sqrt{\lambda_n} \frac{\pi}{6}\right).
\end{equation}

But, since $V=0$ in this region, this result must be consistent with
Eq.~(\ref{eq:phihigh}). This is achieved by the choice
\begin{equation}
A=-\frac{1}{a} \sin\left[\sqrt{\lambda_n}
    \frac{\pi}{2}\right], \label{eq:A}
\end{equation} which ensures
that $\phi_n(\alpha)$ is continuous across the boundary between
``low'' and ``high'' solutions at $\alpha \approx \ell/R$, and the
condition
\begin{equation} \cos\left(\sqrt{\lambda_n} \frac{\pi}{2}
  \right) -\frac{8}{\sqrt{3 \lambda_n}} \sin \left(\sqrt{\lambda_n}
    \frac{\pi}{6} \right)=\sqrt{\frac{2}{\lambda_n}}
  \sin\left(\sqrt{\lambda_n} \frac{\pi}{2}\right) \frac{R}{a},
  \label{eq:Danilov}
\end{equation}
on $\lambda_n$, which ensures that
$\phi_n(\alpha)$ has a continuous first derivative as $\alpha
\rightarrow \ell/R$. We note that if these equations are satisfied
$\lambda_n$, and hence $\phi_n$, is independent of $R$ for $R
\ll |a|$. Indeed, as long as Eqs.~(\ref{eq:Danilov}) and
(\ref{eq:A}) are satisfied the form (\ref{eq:phihigh}) is the result
for $\phi$ for all $\alpha$ such that $\alpha > \ell/R$. Solving
Eq.~(\ref{eq:Danilov}) in the limit $R \ll |a|$ we find the lowest
eigenvalue
\begin{equation} 
\lambda_0=-s_0^2 \left(1 + 1.897
    \frac{R}{a} \right),
\end{equation}
with $s_0=1.00624...$. This is the only negative eigenvalue, and hence
only this channel potential is attractive. So, if we now focus on the
unitary limit, where $|a| \rightarrow \infty$, we have
$\lambda_0=-s_0^2$. Since it can also be shown that the coupling
potentials $P_{nm}$ and $Q_{nm}$ vanish in this regime, the
hyperradial equation (\ref{eq:hyperrad}) in the lowest channel becomes
\begin{equation}
  \frac{\hbar^2}{2 m}\left(-\frac{\partial^2}{\partial R^2} - \frac{s_0^2 +
  \frac{1}{4}}{R^2}\right)f_0(R)=Ef_0(R).
\label{eq:hyperradeqLO}
\end{equation}
This equation will hold for $R \gg \ell$.  If we desire a solution
for negative $E$ the requirement of normalizability for $f_0$ 
mandates that
\begin{equation}
f^{(0)}_0(R)=\sqrt{R}\, K_{is_0} (\sqrt{2} \kappa R),
\label{eq:LOfn}
\end{equation}
where the superscript $(0)$ indicates that we are working in the
unitary limit, while the subscript $0$ refers to the solution for the
hyperchannel corresponding to $\lambda_0$, which is the only one that
supports bound states. The binding energy of these bound states is
related to the $\kappa$ of Eq.~(\ref{eq:LOfn}) by
\begin{equation}
B_t \equiv \frac{\hbar^2 \kappa^2}{m}.
\label{eq:Ekappareln}
\end{equation}
Since the attractive $1/R^2$ potential produces a spectrum that is
unbounded from below some other short-distance physics is needed in
order to stabilize the system. If the two-body potential is known this
short-distance physics is provided by the two-body potential $V$, that
becomes operative for $R \sim \ell$. But an alternative approach is to
add an additional term to Eq.~(\ref{eq:hyperradeqLO}) that summarizes
the impact of the two-body $V$. Here we take this potential to be
a surface delta function at a radius $1/\Lambda$~\cite{Braaten:2004pg}
\begin{equation}
V_{SR}(R)=H_0(\Lambda) \Lambda^2 \delta\left(R - \frac{1}{\Lambda}\right),
\label{eq:VS}
\end{equation}
with $H_0$ adjusted as a function of $\Lambda$ such that the binding
energy of a particular state, say 
$B_t^{(n_*)}$ (with a corresponding
$\kappa_*$, given by (\ref{eq:Ekappareln})), is reproduced. Note that
since $V_{SR}$ is operative only at 
small hyperradii $R \sim 1/\Lambda$ it
corresponds to a three-body force. (See Ref.~\cite{Bedaque:1998kg} for
a realization of this in a momentum-space formalism.)
In physical terms we anticipate $\ell \sim 1/\Lambda$, since we know that
once we consider hyperradii of order $1/\Lambda$ the potential $V$
starts to affect the solutions. 

Given that our focus is on predictions of the theory that are
independent of details of $V$ we can consider the extreme case and
take the limit $\ell \rightarrow 0$. In this limit the form of $K_{is_0}$
as $R \rightarrow 0$ guarantees that once $H_0$ is fixed to give a
bound state at $B_t^{(n_*)}$, the other binding energies in this
hyperradial eigenchannel form a geometric spectrum. Namely,
$B_t^{(n)}=\hbar^2 \kappa_{n}^2/m$ with
\begin{equation}
\kappa_{n}=\left(e^{-\pi/s_0}\right)^{n-n_*} \kappa_*,
\label{eq:Efimov}
\end{equation}
with $n_*$ the index of the bound state corresponding to $\kappa_*$.
Eq.~(\ref{eq:Efimov}) will hold for all $\kappa_n$ such that $\kappa_n
\ll \Lambda$. (Note that now the subscript on $\kappa$ denotes the
index of the bound state in adiabatic channel zero.) The continuous
scale invariance of the $1/R^2$ potential has been broken down to a
discrete scale invariance by the imposition of particular short-distance
physics on the problem through the short-distance potential
(\ref{eq:VS})~\cite{Braaten:2004pg}.

It was subsequently shown that the discrete scale invariance of the
three-body wavefunction which is exact in the limit of infinite
scattering length and zero range also has implications for finite
range. A perturbative calculation of the effect of a finite effective
range on the bound state spectrum showed that the spectrum remains
unchanged \cite{Platter:2008cx} and corrections to binding energies
are of order $(r_0/a)^2$ where $r_0$ is the effective range of the
interaction.

\subsection{Efimov Spectrum}
The hyperspherical methods discussed above can also be used to
obtain the binding energy spectrum at finite scattering length. The
short-distance boundary condition that was used to fix the binding
energy in the unitary limit also determines the bound state spectrum
at finite scattering length. The binding momentum $\kappa_*$
introduced above can therefore be considered as a convenient parameter
that determines the value of all universal few-body observable of the
corresponding universality class.

The exact discrete scaling symmetry observed in the limit of infinite
scattering length also exists if $\kappa_*$ is kept fixed and $a$ and
other variables such as the energy are rescaled
\begin{eqnarray}
  \label{eq:4}
  \kappa_*\rightarrow
  \kappa_*~,\hspace{5mm}a\rightarrow\mathcal{S}_0^{m}~a~,
\hspace{5mm}E\rightarrow\mathcal{S}_0^{-2m}E~.
\end{eqnarray}
Observables such as binding energy and cross sections scale with
integer powers of $\mathcal{S}_0=\exp(\pi/s_0)$ under this symmetry.  For example,
the binding energy of an Efimov trimer which is a function of $a$ and
$\kappa_*$ scales
\begin{equation}
  \label{eq:1}
  B_t^{(n)}(\mathcal{S}_0^m a,\kappa_*)=\mathcal{S}_0^{-2m}B_t^{(n-m)}(a,\kappa_*)~.
\end{equation}
This implies for positive scattering length
\begin{equation}
  \label{eq:2}
  B_t^{(n)}(a,\kappa_*)=F_n (2 s_0\ln(a \kappa_*))\frac{\hbar^2\kappa_*^2}{m}~.
\end{equation}
The function $F_n$ parametrizes the scattering length dependence of
all Efimov trimers exactly in the limit of vanishing range. The
function $F_n$ satisfies
\begin{equation}
  \label{eq:3}
  F_n(x+2m\pi)=(e^{-2\pi/s_0})^mF_{n-m}(x)~.
\end{equation}
\begin{figure}[t!]
\centerline{\includegraphics*[width=8cm,angle=0]{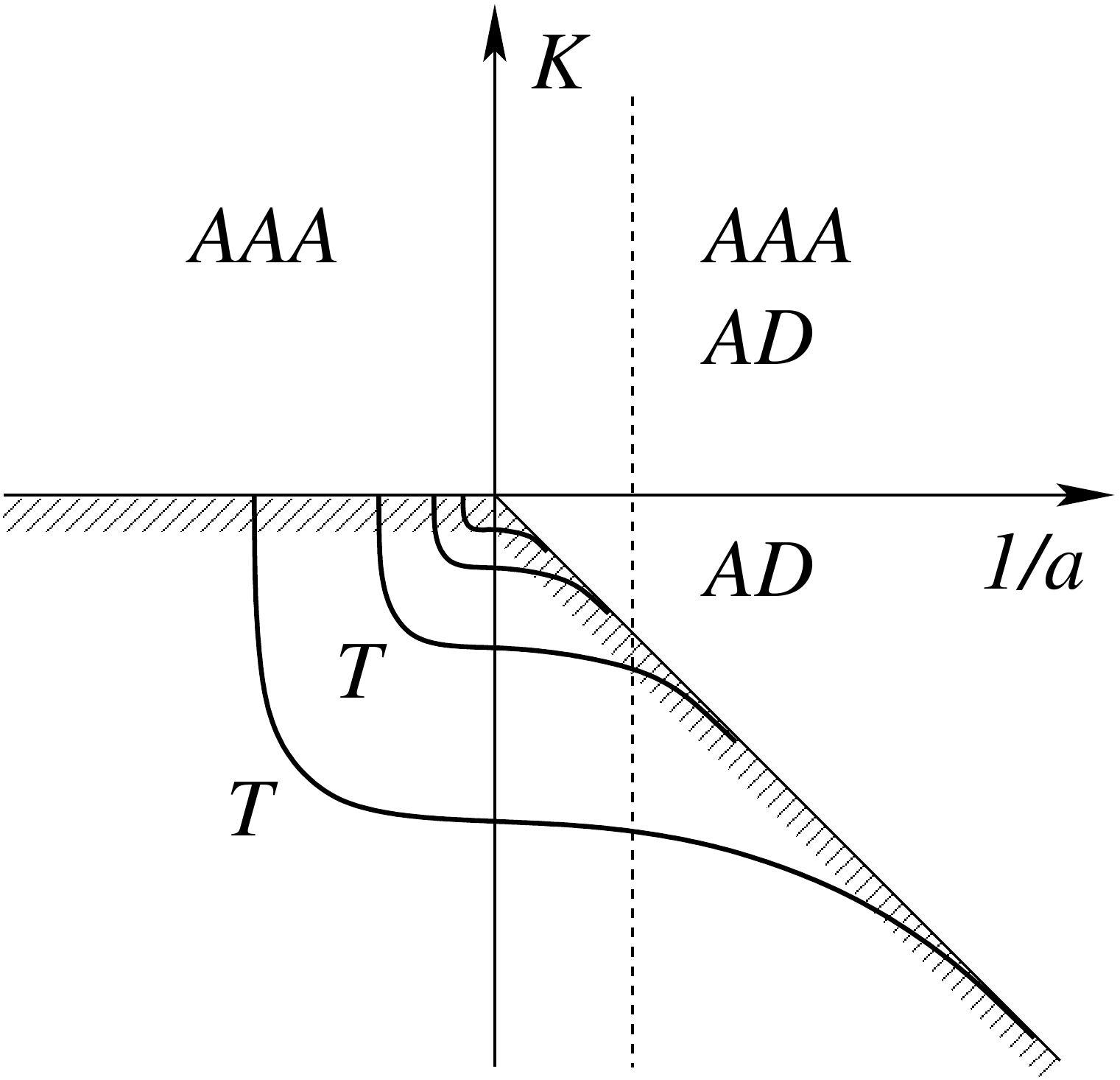}}
\caption
{
\baselineskip 12pt  
The Efimov plot for the three-body problem. We show
$K\equiv {\rm sgn}(E)(m|E|)^{1/2}/\hbar$ versus the inverse scattering 
length. The allowed regions for
three-atom scattering states and atom-dimer scattering states are
labeled $AAA$ and $AD$, respectively. The heavy lines labeled $T$
are two of the infinitely many branches of Efimov states.
The cross-hatching indicates the threshold for scattering states.
States along the vertical dashed line have a fixed scattering length.}
\label{fig:efimovplot}
\end{figure}
The scattering length dependence of the bound state spectrum is shown
in Fig. \ref{fig:efimovplot}. We plot the quantity $K\equiv {\rm
  sgn}(E)(m|E|)^{1/2}/\hbar$ against the inverse scattering length.
For bound states $K$ corresponds to the binding momentum.  The lines
denote Efimov trimers below the threshold. The threshold for
scattering states is denoted by the hatched area. Only a
few of the infinitely many Efimov branches are shown. A given physical
system has a fixed scattering length value and is denoted by the
vertical dashed line. Changing the parameter $\kappa_*$ by a factor
$\mathcal{S}_0$ corresponds to multiplying each branch of trimers with
this factor without changing their shapes.  One important result is
that three-bound states exist for positive and negative scattering
length. This is remarkable for the latter case since the two-body
subsystem is unbound for $a<0$. At a negative scattering length that
we denote with $a'_*$, a bound state with a give $\kappa_*$ has zero
binding energy. As the scattering length is increased the binding
energy gets larger until it crosses the atom-dimer threshold at the
positive scattering length $a_*$. The quantities $a_*$ and $a'_*$ can
also be used to quantify a universality class of Efimov states.

\subsection{Universal Properties}
\label{sec:universal-properties}
Other calculable observables will also display the discrete scaling
symmetry that we just discussed for the bound state spectrum. The
atom-dimer cross section fulfils for example the constraint
\begin{eqnarray}
  \label{eq:5}
  \sigma_{\rm AD}(\mathcal{S}_0^{-2m} E; \mathcal{S}_0^m a,\kappa_*)
=\mathcal{S}_0^{2m} \sigma_{AD}(E;a,\kappa_*)~
\end{eqnarray}
under rescaling. At $E=0$ the cross section is related to the
atom-dimer scattering length $\sigma_{\rm AD}=4\pi|a_{\rm AD}|^2$.
This implies that the atom-dimer scattering length can be written as
\begin{equation}
  \label{eq:6}
  a_{\rm AD}=f(2s_0\ln(a \kappa_*)) a~,
\end{equation}
where $f(x)$ is periodic function with period $2\pi$.

An observable that has been crucial for the experimental detection of
Efimov physics in ultracold atoms is the three-body recombination
rate. In ultracold gases, atoms can undergo inelastic three-body
collisions in which a two-body bound state is formed. The dimer and
remaining atom gain kinetic energy in this process and can leave the
atomic trap. Such processes lead to a measurable loss of particles in
the atomic trap. For negative scattering length, atoms can only recombine
into deep dimers that have binding energy of order
$\hbar^2/(m R^2)$. For positive
scattering length, atoms can recombine into shallow dimers with
binding energy $\hbar^2/(m a^2)$ and deep dimers. The recombination rate
that is a measure for the loss rate of atoms scales as $\hbar a^4/m$ times
a log-periodic coefficient that is a function of $a$ and $\kappa_*$. The
recombination rate constant $\alpha$ can therefore be written as
\begin{equation}
  \label{eq:7}
  \alpha(a)=g(2 s_0 \ln(a \kappa_*)) \frac{\hbar a^4}{m}\,.
\end{equation}
\begin{figure}[tb!]
\centerline{\includegraphics*[width=10cm,angle=0]{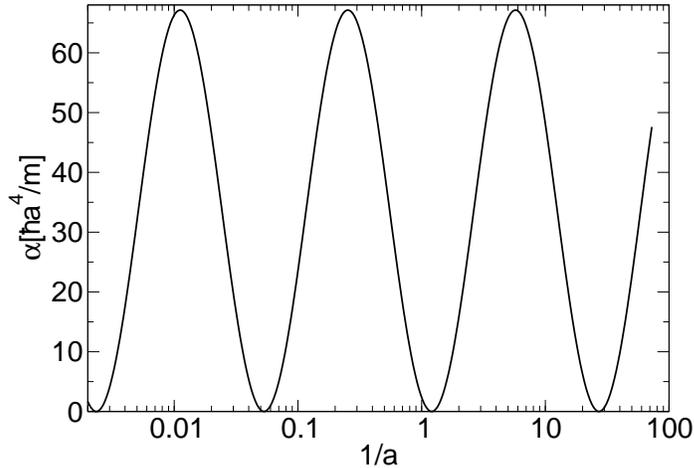}}
\caption{\label{fig:reco}
\baselineskip 12pt  
The recombination rate constant $\alpha$
for positive scattering length $a$ in units of $\hbar a^4/m$ as 
a function of $1/a$ (in arbitrary units). 
}
\end{figure}
The analytic form of the function $g(x)$ is known and can be found in
\cite{Braaten:2004rn,Braaten:2006vd}. Here we want to focus on the qualitative
features as shown in Fig. \ref{fig:reco} for positive scattering
length. At positive scattering length interference effects lead to
log-periodically spaced minima in the recombination rate.  At
negative scattering length free atoms can only recombine into deep
dimers. This process will be enhanced dramatically whenever an Efimov
trimer is at threshold.

A different perspective on Efimov physics can be gained by keeping the
two-body scattering length fixed and varying the three-body parameter.
As a result all three-body observables are correlated and will lead to
correlation lines when plotted against each other. One of them
(well-known from nuclear physics) is the Phillips line
\cite{Phillips68}, the
correlation that results when the trimer binding energy is plotted
against the atom-dimer scattering length. Such correlations had been
observed frequently in nuclear physics, where different phase shift
equivalent potentials were employed in few-body calculations.

Since one three-body parameter is required for a description of the
three-body system with zero-range interactions it is natural to ask
how many parameters are needed for calculations in the $n$-body
system. A first step towards answering this question was performed in
Ref.~\cite{Platter:2004qn}. The authors of this work showed that the
two-body scattering length and one three-body parameter are sufficient
to make predictions for four-body observables. Results in a more
detailed analysis \cite{Hammer:2006ct} also lead to the conclusion
that every trimer state is tied to two universal tetramer states with
binding energies related to the binding energy of the next shallower
trimer. In the unitary limit $1/a=0$, the relation between the binding
energies was found as:
\begin{equation}
  \label{eq:4body1}
  B_4^ {(0)}\approx 5\, B_t\quad {\rm and} \quad  B_4^{(1)}\approx 1.01\, B_t~,
\end{equation}
where $B_4^ {(0)}$ denotes the binding energy of the deeper of the two tetramer
states and $B_4^ {(1)}$ the shallower of the two.

A recent calculation by von Stecher, d'Incao and Greene \cite{vStech08}
supports these findings and extends them to higher numerical
accuracy.
For the relation between universal three- and four-body
bound states in the unitary limit, they found
 \begin{equation}
  \label{eq:4body2}
 B_4^ {(0)}\approx 4.57\, B_t\quad {\rm and} \quad  B_4^{(1)}\approx 1.01\, B_t~,
\end{equation}
which is consistent with the results given in Eq.~(\ref{eq:4body1})
within the numerical accuracy.

\begin{figure}[tb!]
\centerline{\includegraphics*[clip=true,width=10cm,angle=0]{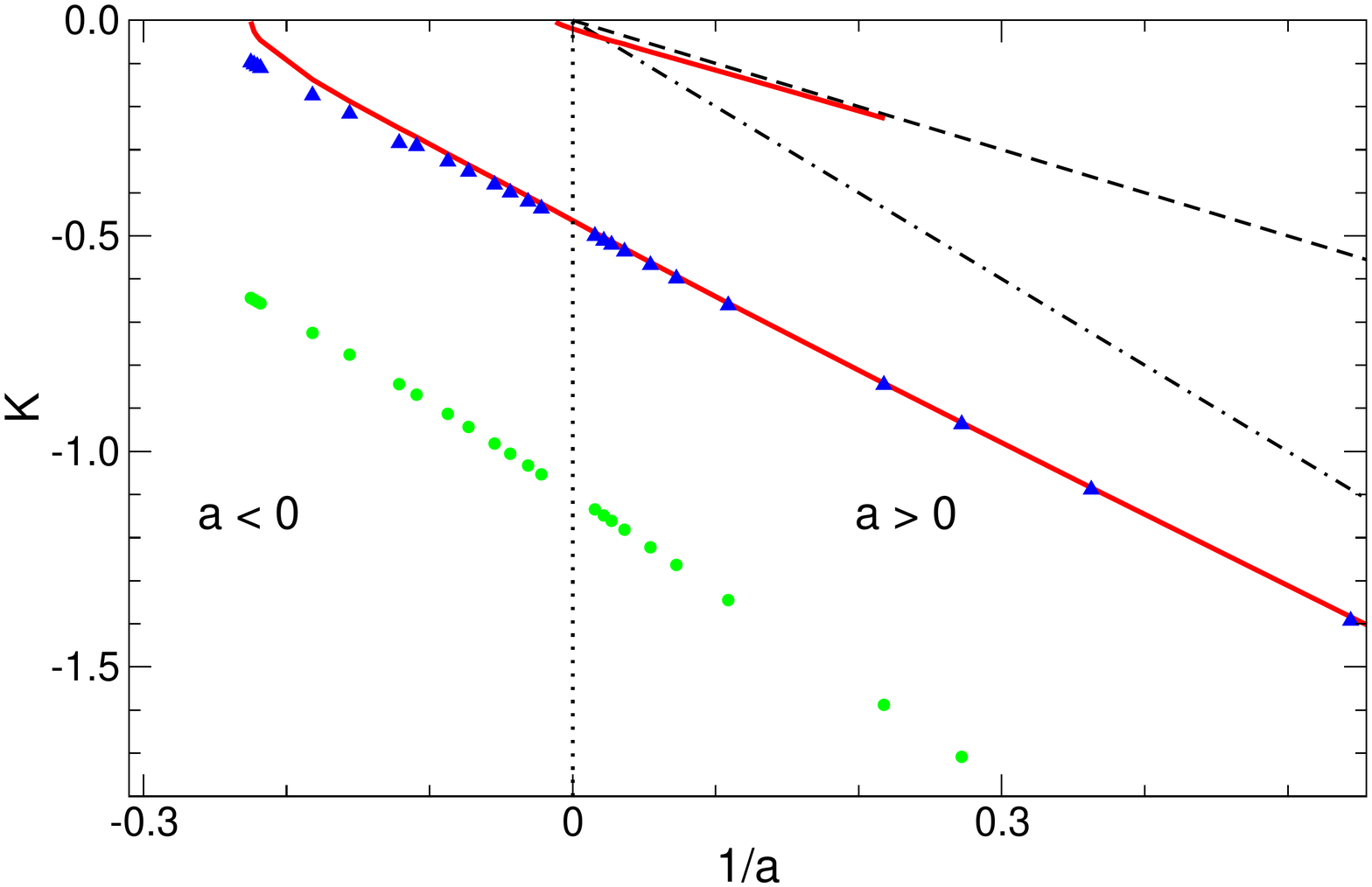}}
\caption{\label{fig:efimov-4body} 
\baselineskip 12pt 
The extended Efimov plot for the four-body problem. We show
$K\equiv {\rm sgn}(E)(m|E|)^{1/2}/\hbar$ versus the inverse scattering 
length. Both quantities are given in arbitrary units.
The circles and triangles indicate the four-body ground and excited
states, respectively, while the lower (upper) solid
lines give the thresholds for decay into a ground state (excited state)
trimer and a particle. The dash-dotted (dashed) lines give the thresholds
for decay into two dimers (a dimer and two particles).  The vertical dotted
line indicates infinite scattering length.  }
\end{figure}

The results obtained by the Hammer and Platter in Ref.~\cite{Hammer:2006ct}
were furthermore presented in the form of an extended Efimov plot, shown in
Fig.~\ref{fig:efimov-4body}. Four-body states must have a binding energy
larger than the one of the deepest trimer state. The corresponding threshold
is denoted by lower solid line in Fig.~~\ref{fig:efimov-4body}.  The threshold
for decay into the shallowest trimer state and an atom is indicated by the
upper solid line. At positive scattering length, there are also scattering
thresholds for scattering of two dimers and scattering of a dimer 
and two particles
indicated by the dash-dotted and dashed lines, respectively.  The vertical
dotted line denotes infinite scattering length. 

An extended version of this four-body Efimov plot was also presented
by von Stecher, d'Incao and Greene in Ref.~\cite{vStech08}.  They
calculated more states with higher numerical accuracy and extended the
calculation of the four-body states to the thresholds where they
become unstable.
From these results they extracted the negative values of the 
scattering lengths at which the binding energies of the tetramer 
states become zero and found
\begin{equation}
\label{eq:tetra-scatteringlengths}
  a^*_{4,0}\approx 0.43 a'_*\quad{\rm and}\quad a^*_{4,1}\approx 0.92 a'_*~.
\end{equation}
These numbers uniquely specify the relative position of three- and four-body
recombination resonances. This was
the key information for the subsequent observation 
of these states in ultracold atoms by Ferlaino et al.
\cite{Ferlaino:2009zz}.

Calculations for larger number of particles using a model that
incorporates the universal behavior of the three-body system were
carried our by von Stecher \cite{stecher:2009b}. His findings indicate
that there is at least one $N$-body state tied to each Efimov
trimer and numerical evidence was also found for a second excited
5-body state.

\subsection{Observation in Ultracold Atoms}
\label{sec:observ-ultr-atoms}
The first experimental evidence for Efimov physics in ultracold atoms
was presented by Kraemer et al.\ \cite{Kraemer:2006} in 2006. This
group used $^{133}$Cs atoms in the lowest hyperfine spin state. They
observed resonant enhancement of the loss of atoms from three-body
recombination that can be attributed to an Efimov trimer crossing the
three-atom threshold. 
Kraemer et al.\ also observed a minimum
in the three-body recombination rate that can be interpreted as an
interference effect associated with Efimov physics. In a subsequent
experiment with a mixture of $^{133}$Cs atoms and dimers, Knoop et
al.\ observed a resonant enhancement in the loss of atoms and dimers
\cite{Knoop:2008}. This loss feature can be explained by an Efimov
trimer crossing the atom-dimer threshold \cite{Helfrich:2009uy}. The
most exciting recent developments in the field of Efimov physics
involve universal tetramer states. Ferlaino et al.\ 
observed two tetramers in an ultracold gas of $^{133}$Cs atoms
\cite{Ferlaino:2009zz} that confirm the results by Hammer and
Platter~\cite{Platter:2004qn,Hammer:2006ct} and von Stecher, D'Incao
and Greene~\cite{vStech08}.

Recent experiments with other bosonic atoms have provided even
stronger evidence of Efimov physics in the three- and four-body
sectors. Zaccanti et al.\ measured the three-body recombination rate and
the atom-dimer loss rate in a ultracold gas of $^{39}$K atoms
\cite{Zaccanti:2008}. They observed two atom-dimer loss resonances and
two minima in the three-body recombination rate at large positive
values of the scattering length.  The positions of the loss features
are consistent with the universal predictions with discrete scaling
factor 22.7. They also observed loss features at large negative
scattering lengths. Barontini et al.\ obtained the first evidence of
the Efimov effect in a heteronuclear mixture of $^{41}$K and $^{87}$Rb
atoms \cite{Barontini:2009}. They observed 3-atom loss resonances at
large negative scattering lengths in both the K-Rb-Rb and K-K-Rb
channels, for which the discrete scaling factors are 131 and $3.51
\times 10^5$, respectively. Gross {\it et al.} measured the three-body
recombination rate in an ultracold system of $^7$Li atoms
\cite{Gross:2009}. They observed a 3-atom loss resonance at a large
negative scattering length and a three-body recombination minimum at a
large positive scattering length. The positions of the loss features,
which are in the same universal region on different sides of a
Feshbach resonance, are consistent with the universal predictions with
discrete scaling factor 22.7. Pollack {\it et al.} measured the
three-body recombination in a
system of $^7$Li atoms in a hyperfine state different from the system
considered by Gross {\it et al.} \cite{pollack:2009}. They observed a total of 11 three-
and four-body loss features. The features obey the
universal relations on each side of the Feshbach resonance separately,
however, a systematic error of \~ 50 \% is found when features on different 
sides of the Feshbach resonance are compared.

Efimov physics has also been observed in three-component systems of
$^6$Li atom. For the three lowest hyperfine states of $^6$Li atoms,
the three pair scattering lengths approach a common large negative
value at large magnetic fields and all three have nearby Feshbach
resonances at lower fields that can be used to vary the scattering
lengths \cite{Bartenstein:2005}.  The first experimental studies of
many-body systems of $^6$Li atoms in the three lowest hyperfine states
have recently been carried out by Ottenstein et al.\
\cite{Ottenstein:2008} and by Huckans et al.\ \cite{Huckans:2008fq}.
Their measurements of the three-body recombination rate revealed a narrow
loss feature and a broad loss feature in a region of low magnetic
field.  Theoretical calculations of the three-body recombination rate
supported the interpretation that the narrow loss feature arises from
an Efimov trimer crossing the 3-atom threshold
\cite{Braaten:2008wd,NU:2009,Schmidt:2008fz}.  Very recently, another
narrow loss feature was discovered in a much higher region of the
magnetic field by Williams et al.\ \cite{Williams:2009} and by Jochim
and coworkers. Williams et al.\ used measurements of the three-body
recombination rate in this region to determine the complex three-body
parameter that governs Efimov physics in this system.  This parameter,
together with the three scattering lengths as functions of the
magnetic field, determine the universal predictions for $^6$Li atoms
in this region of the magnetic field.
\section{Applications in Nuclear Physics}
\label{sec:nucphys}
The properties of hadrons and nuclei are determined by 
quantum chromodynamics (QCD), a non-abelian gauge theory formulated 
in terms of quark and gluon degrees of freedom. At low energies, 
however, the appropriate degrees of freedom are the hadrons. 
Efimov physics and the unitary limit can serve as a useful starting 
point for effective field theories (EFTs) 
describing hadrons and nuclei at very low energies.
For convenience, we will now work in natural units where $\hbar=c=1$.

In nuclear physics, there are a number of EFTs which are all useful
for a certain range of systems
\cite{Beane:2000fx,Bedaque:2002mn,Epelbaum:2008ga}.  At very low
energies, where Efimov physics plays a role, all interactions can be
considered short-range and even the pions can be integrated out. This
so-called \lq\lq pionless EFT'' is formulated in an expansion of the
low-momentum scale $\Mlo$ over the high-momentum scale $\Mhi$. It can
be understood as an expansion around the limit of infinite scattering
length or equivalently around threshold bound states. Its breakdown
scale is set by one-pion exchange, $\Mhi\sim M_\pi$, while $\Mlo \sim
1/a \sim k$.  For momenta $k$ of the order of the pion mass $M_\pi$,
pion exchange becomes a long-range interaction and has to be treated
explicitly.  This leads to the chiral EFT whose breakdown scale $\Mhi$
is set by the chiral symmetry breaking scale $\Lambda_\chi$. The
pionless theory relies only on the large scattering length and is
independent of the short-distance mechanism generating it. This theory
is therefore ideally suited to unravel universal phenomena driven by
the large scattering length such as limit cycle physics
\cite{Mohr:2005pv,Braaten:2003eu} and the Efimov effect
\cite{Efimov-70}.  In this section, we will focus on the
aspects of nuclear effective field theories related to Efimov
physics. For more complete overviews of the application of effective
field theories to nuclear phenomena in general, a number of excellent
reviews are available
\cite{Beane:2000fx,Bedaque:2002mn,Epelbaum:2005pn,Epelbaum:2008ga}.

\subsection{Few-Nucleon System}
\label{sec:few-nucleon-system}
In the two-nucleon system, the pionless theory reproduces the well
known effective range expansion in the large scattering length limit.
The renormalized S-wave scattering amplitude to next-to-leading order
in a given channel takes the form \beqa T_2 (k) &=& \frac{4\pi}{m}
\frac{1}{-1/a-ik} \left[ 1-\frac{r_0 k^2/2}{-1/a-ik}+\ldots \right]\,,
\eeqa where $k$ is the relative momentum of the nucleons and the dots
indicate corrections of order $(\Mlo /\Mhi)^2$ for typical momenta
$k\sim\Mlo$. In the language of the renormalization group, this
corresponds to an expansion around the non-trivial fixed point for
$1/a=0$ \cite{Kaplan:1998we,Birse:1998dk}. The pionless EFT becomes
very useful in the two-nucleon sector when external currents are
considered and has been applied to a variety of electroweak processes.
These calculations are reviewed in detail in
Refs.~\cite{Beane:2000fx,Bedaque:2002mn}.

Here we focus on the three-nucleon system.
It is convenient (but not mandatory) to write the 
theory using so-called \lq\lq dimeron'' 
auxiliary fields \cite{Kaplan:1996nv}.
We need two dimeron fields, one for each S-wave channel:
(i) a field $t_i$ with
spin (isospin) 1 (0) representing two nucleons interacting in the $^3 S_1$
channel (the deuteron) and
(ii) a field $s_a$ with
spin (isospin) 0 (1) representing two nucleons interacting in the $^1 S_0$
channel \cite{Bedaque:1999ve}:
\beqa
{\cal L}&=&N^\dagger \Big(i\partial_t +\frac{\vec{\nabla}^2}
{2m}\Big)N - t^\dagger_i \Big(i\partial_t -\frac{\vec{\nabla}^2}
{4m} -\Delta_t \Big) t_i \nonumber\\
&-& s^\dagger_a \Big(i\partial_t -\frac{\vec{\nabla}^2}
{4m} -\Delta_s \Big) s_a
- \frac{g_t}{2}\Big( t^\dagger_i N^T \tau_2 \sigma_i
\sigma_2 N +h.c.\Big) 
\nonumber \\
&-&\frac{g_s}{2}\Big(s^\dagger_a N^T \sigma_2 \tau_a \tau_2 N +h.c.\Big) 
-G_3 N^\dagger \Big[ g_t^2
(t_i \sigma_i)^\dagger (t_j\sigma_j) \nonumber \\
&+& \frac{g_t g_s}{3} \left( (t_i\sigma_i)^\dagger
(s_a \tau_a) + h.c. \right) 
+ g_s^2 (s_a \tau_a)^\dagger (s_b \tau_b) \Big] N +\ldots\,,
\label{lagd}
\eeqa
where $i,j$ are spin and $a,b$ are isospin indices
while  $g_t$, $g_s$, $\Delta_t$, $\Delta_s$
and $G_3$ are the bare coupling constants.
The Pauli matrices
$\sigma_i$ $(\tau_a)$ operate in spin (isospin) space, respectively.
This Lagrangian goes beyond leading order and
already includes the effective range terms. The coupling
constants $g_t$, $\Delta_t$, $g_s$, $\Delta_s$ are matched to the
scattering lengths $a_\alpha$ and effective ranges $r_{0\alpha}$
in the two channels ($\alpha=s,t$). Alternatively, one can 
match to the position of the 
bound state/virtual state pole $\gamma_\alpha$ 
in the $T$-matrix instead of the 
scattering length which often improves convergence \cite{Phillips:1999hh}.

The term proportional to $G_3$ constitutes a Wigner-$SU(4)$ symmetric
three-body interaction. It only contributes in the spin-doublet S-wave
channel.  When the auxiliary dimeron fields $t_i$ and $s_a$ are
integrated out, an equivalent form containing only nucleon fields is
obtained.  At leading order when the effective range corrections are
neglected, the spatial and time derivatives acting on the dimeron
fields are omitted and the field is static.  The coupling constants
$g_\alpha$ and $\Delta_\alpha$, $\alpha=s,t$ are then not independent
and only the combination $g_\alpha^2/\Delta_\alpha$ enters in
observables. This combination can then be matched to the scattering
length or pole position.

\begin{figure}[tb!]
\bigskip
\centerline{\includegraphics*[width=8cm,angle=0]{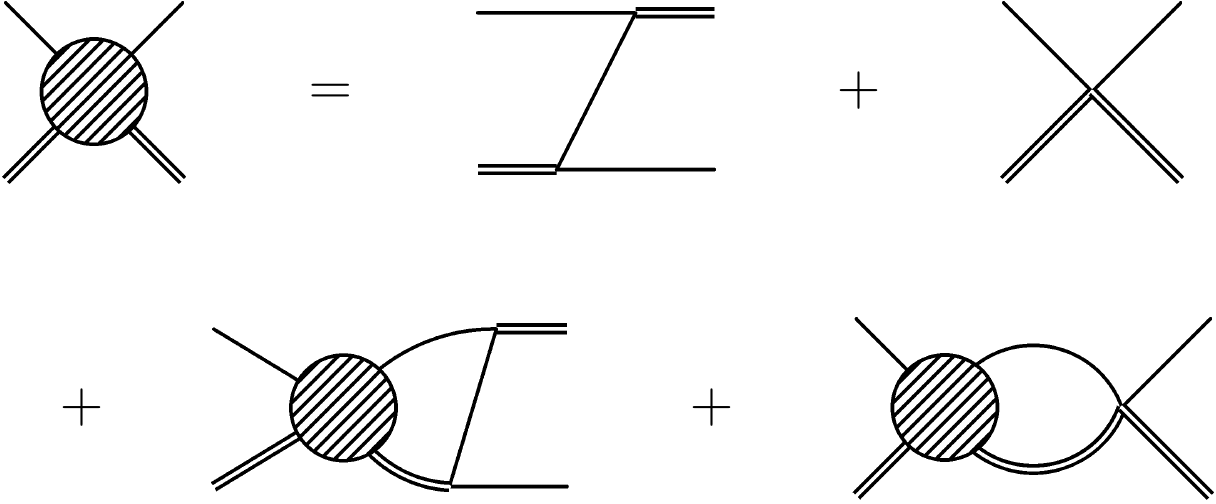}}
\caption{
\baselineskip 12pt  
The integral equation for the particle-dimeron scattering amplitude.
The single (double) line indicates the particle (dimeron)
propagator.}
\label{fig:ineq}
\end{figure}

The simplest three-body process to consider is neutron-deuteron
scattering below the breakup threshold.  In order to focus on the main
aspects of renormalization, we suppress all spin-isospin indices and
complications from coupled channels in the three-nucleon problem.
This is equivalent to a system of three spinless bosons with large
scattering length. If the scattering length is positive, the
particles form a two-body bound state analog to the deuteron which we
generically call dimeron. The leading order integral equation for
particle-dimeron scattering is shown schematically in
Fig.~\ref{fig:ineq}.
Projected on total orbital angular momentum $L=0$, it takes the following form:
\beqa
T_3(k, p; \, E) &=& \frac{16}{3 a} M (k, p; \, E) + \frac{4}{\pi} 
\int_0^\Lambda dq\, q^2\, 
T_3(k, q; \, E) \nonumber \\
&&\times \, \frac{ M (q, p; \, E)}{- 1/a + \sqrt{3 q^2/4 - m E - i \epsilon}} 
\,, 
\label{STM}
\eeqa
where the inhomogeneous term reads
\beq 
M (k, p; \, E) =
\frac{1}{2 k p} \ln \left( \frac{k^2 + k p + p^2 - m E}{k^2 - k p +
    p^2 - m E} \right) + \frac{H (\Lambda )}{\Lambda^2}\,. 
\eeq Here,
$H(\Lambda)$ is a running coupling constant that determines the
strength of the three-body force $G_3(\Lambda)=2mH(\Lambda)/\Lambda^2$
and $\Lambda$ is a UV cutoff introduced to regularize the integral
equation.  Note that the three-body force is enhanced and enters
already at leading order in this theory.  The magnitude of the
incoming (outgoing) relative momenta is $k$ ($p$) and $E = 3 k^2/(4 m)
- 1/(ma^2)$.  The on-shell point corresponds to $k = p$ and the phase
shift can be obtained via $k \cot \delta = 1/T_3(k, k; \, E)+ik$.  For
$H \equiv 0$ and $\Lambda \to \infty$, Eq.~(\ref{STM}) reduces to the
STM equation of Skorniakov and Ter-Martirosian
\cite{Skorniakov:1957aa}. It is well known that the STM equation has
no unique solution \cite{Danilov:1961aa}.  The regularized STM
equation has a unique solution for any given (finite) value of the
ultraviolet cutoff $\Lambda$ but the solution strongly depends on the
value of $\Lambda$.  In the EFT framework, cutoff independence of the
amplitude is achieved by an appropriate ``running'' of $H (\Lambda )$
\cite{Bedaque:1998kg,Bedaque:1998km}:
\beq
H(\Lambda)=\frac{\cos[s_0
  \ln(\Lambda/\Lambda_*)+\arctan s_0]} {\cos[s_0
  \ln(\Lambda/\Lambda_*)-\arctan s_0]}\,,
\label{eq:Heq}
\eeq
where $\Lambda_*$ is a dimensionful three-body parameter generated
by dimensional transmutation. 
\begin{figure}[ht!]
\bigskip
\centerline{\includegraphics*[width=8cm,angle=0]{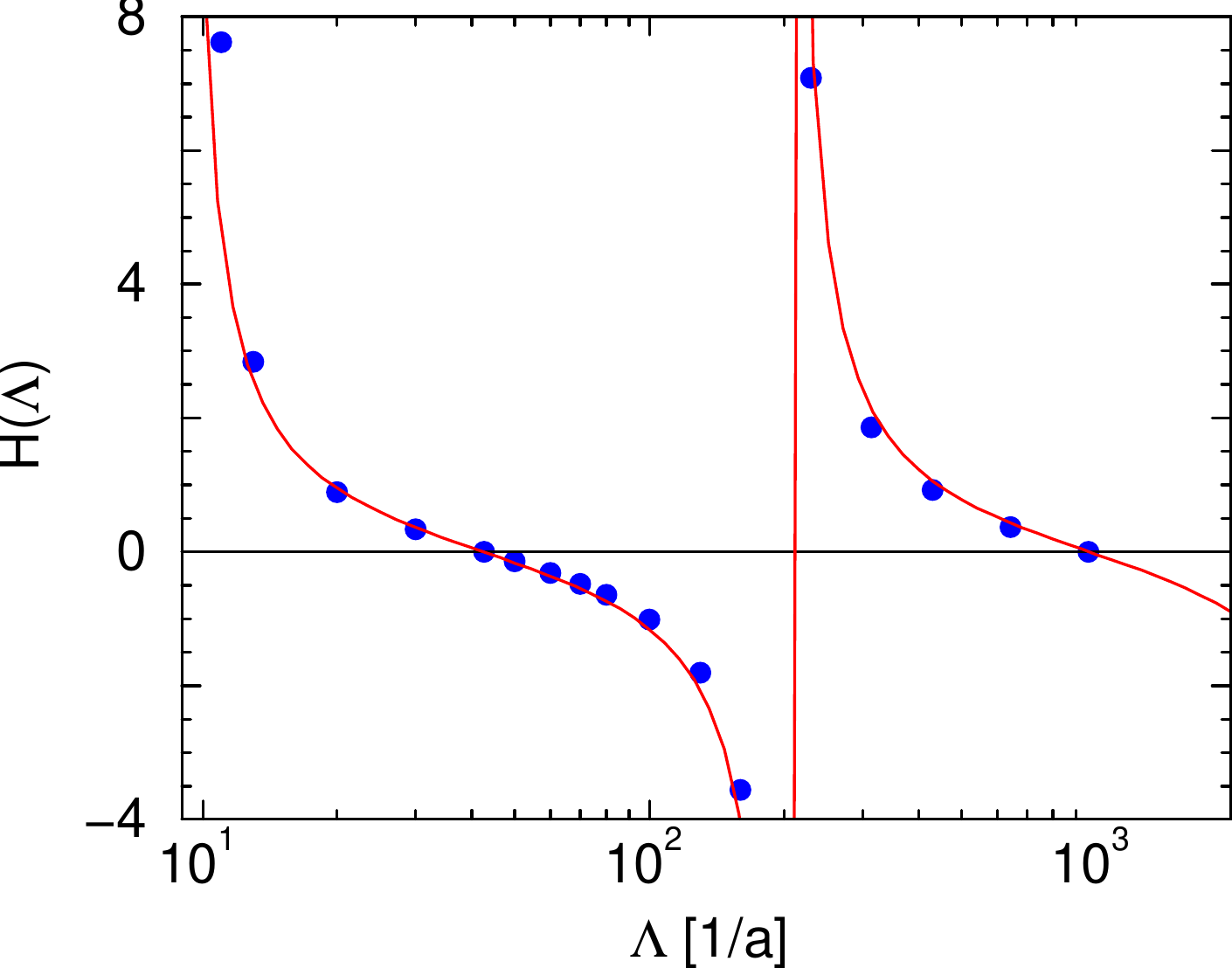}
}
\caption{
\baselineskip 12pt 
The three-body coupling $H$ as a function of the cutoff $\Lambda$
for a fixed value of the three-body parameter $\Lambda_*$.
The solid line shows the analytical expression (\ref{eq:Heq}), while
the dots show results from the numerical solution of Eq.~(\ref{STM}).
}
\label{fig:limit}
\end{figure}
The dependence of the three-body coupling $H$ on the cutoff $\Lambda$
is shown in Fig.~\ref{fig:limit} for a fixed value of the three-body
parameter $\Lambda_*$.  The solid line shows the analytical expression
(\ref{eq:Heq}), while the dots show results from the numerical
solution of Eq.~(\ref{STM}).  A good agreement between both methods is
observed, indicating that the renormalization is well under
control. Adjusting $\Lambda_*$ to a single three-body observable
allows to determine all other low-energy properties of the three-body
system.\footnote{Note that the choice of the three-body parameter
  $\Lambda_*$ is not unique, for alternative definitions see
  \cite{Braaten:2004rn}.}  Because $H(\Lambda)$ in Eq.~(\ref{eq:Heq})
vanishes for certain values of the cutoff $\Lambda$ it is possible to
eliminate the explicit three-body force from the equations by working
with a fixed cutoff that encodes the dependence on $\Lambda_*$. This
justifies tuning the cutoff $\Lambda$ in the STM equation to reproduce
a three-body datum and using the same cutoff to calculate other
observables as suggested by Kharchenko \cite{Kharchenko:1973aa}.
Equivalently, a subtraction can be performed in the integral
equation~\cite{Hammer:2000nf,Afnan:2003bs}.  In all cases one
three-body input parameter is needed for the calculation of
observables.

The discrete scaling symmetry of the Efimov spectrum is manifest 
in the running of  the coupling $H (\Lambda )$.
The spectrum of three-body bound states of this EFT is exactly the Efimov
spectrum. 
The integral equations for the three-nucleon problem derived from the
Lagrangian (\ref{lagd}) are a generalization of Eq.~(\ref{STM}).
(For their explicit form and derivation, 
see e.g. Ref.~\cite{Bedaque:2002yg}.)

For S-wave nucleon-deuteron scattering in the spin-quartet channel
only the spin-1 dimeron field contributes and the integral equation becomes
\cite{Skorniakov:1957aa,Bedaque:1997qi,Bedaque:1998mb}
\begin{equation}
  \label{eq:quartet}
  T_3^{(3/2)}(p,k;E)=-\frac{4\gamma_t}{3}K(p,k)
      -\frac{1}{\pi}\int_0^\infty dq\,
  q^2 D_t(q;E)\, K(p,q)\, T_3^{(3/2)}(q,k;E)~,
\end{equation}
where
\begin{equation}
  K(p,k)=\frac{1}{p k}\ln\left(\frac{p^2+p k+k^2-mE}{p^2-pk+k^2-mE}\right)~,
\end{equation}
$D_t(q;E)$ is the full spin-1 dimeron propagator and 
$\gamma_t \approx 45$~MeV the deuteron pole momentum.
This integral equation has a unique solution for $\Lambda \to \infty$ 
and there is no three-body force in the first few orders. 
An S-wave three-body force is forbidden by the Pauli principle 
in this channel since all nucleon spins have to be aligned to obtain
$J=3/2$. The spin-quartet scattering phases 
$k \cot \delta^{(3/2)} = 1/T_3^{(3/2)}(k, k; \, E)+ik$
can therefore be predicted to high precision from two-body data alone.

In the spin-doublet channel both dimeron fields as well as the
three-body force in the Lagrangian (\ref{lagd}) contribute
\cite{Bedaque:1999ve}. This leads to a pair of coupled integral
equations for the T-matrix.  The renormalization of this equation is
easily understood in the unitary limit which corresponds to a Wigner
$SU(4)$ symmetry of the theory \cite{Wigner37}.  In the unitary limit
these two integral equations decouple.  One of the two equations has
the same structure as the equation for the bosonic problem
(\ref{STM}), while the other one is similar to the equation in the
quartet channel (\ref{eq:quartet}).  Thus, one needs one new parameter
which is not determined in the 2N system in order to fix the (leading)
low-energy behavior of the 3N system in this channel. This parameter
corresponds to the $SU(4)$ symmetric three-body force 
proportional to $G_3$ in the Lagrangian (\ref{lagd})
\cite{Bedaque:1999ve}. The three-body parameter gives a natural
explanation of universal correlations between different three-body
observables such as the Phillips line, a correlation between the
triton binding energy and the spin-doublet neutron-deuteron scattering
length \cite{Phillips68}.  These correlations are purely driven by the
large scattering length independent of the mechanism responsible for
it.  If the spin-doublet neutron-deuteron scattering length is given,
the triton binding energy is predicted. In this scenario
the triton emerges as
an Efimov state. The scenario can be tested by using the effective theory to
predict other three-body observables.

\begin{figure}[tb!]
\centerline{\includegraphics*[width=8cm,angle=0,clip=true]{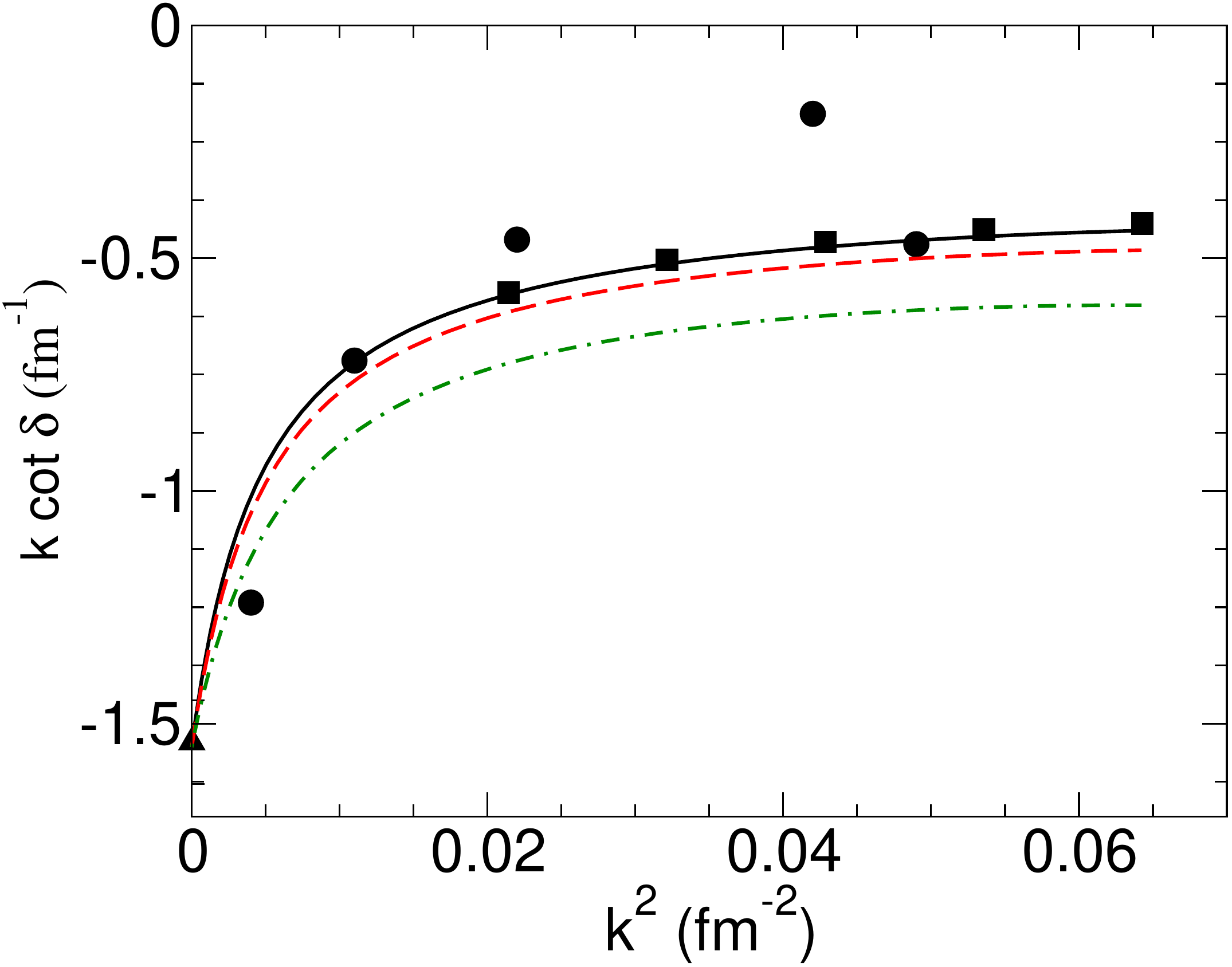}}
\caption{
\baselineskip 12pt 
Phase shifts for neutron-deuteron scattering below the deuteron
breakup at LO (dash-dotted line), NLO (dashed line), and N$^2$LO
(solid line). The filled squares and circles give the results of
a phase shift analysis and a calculation using AV18 and the 
Urbana IX three-body force, respectively.
}
\label{fig:nddoublet}
\end{figure}

Higher-order corrections to the amplitude including the ones due to 2N
effective range terms can be included perturbatively.  This was first
done at NLO for the scattering length and triton binding energy in
\cite{Efimov:1991aa} and for the energy dependence of the phase shifts
in \cite{Hammer:2000nf}.  In
Refs.~\cite{Bedaque:2002yg,Griesshammer:2004pe}, it was demonstrated
that it is convenient to iterate certain higher order range terms in
order to extend the calculation to N$^2$LO. Here, also a subleading
three-body force was included as required by dimensional
analysis. More recently, Platter and Phillips showed using the
subtractive renormalization that the leading three-body force is
sufficient to achieve cutoff independence up to N$^2$LO in the
expansion in $\Mlo/\Mhi$ \cite{Platter:2006ev}. The results for the
spin-doublet neutron-deuteron scattering phase shift at LO
\cite{Bedaque:1999ve}, NLO \cite{Hammer:2000nf}, and N$^2$LO
\cite{Platter:2006ad} are shown in Fig.~\ref{fig:nddoublet}.
There is excellent agreement with the available phase shift analysis
and a calculation using a phenomenological NN interaction.  
From dimensional analysis, one would expect the subleading 
three-body force at N$^2$LO. Whether
there is a suppression of the subleading three-body force or simply a
correlation between the leading and subleading contributions is not
understood.

Three-nucleon channels with higher orbital angular momentum are
similar to the spin-quartet for S-waves and three-body forces do not
appear until very high orders \cite{Gabbiani:1999yv}.  A general
counting scheme for three-body forces based on the asymptotic behavior
of the solutions of the leading order STM equation was proposed 
by Grie\ss hammer
\cite{Griesshammer:2005ga}.  A complementary approach to the
few-nucleon problem is given by the renormalization group where the
power counting is determined from the scaling of operators under the
renormalization group transformation \cite{Wilson-83}. This method
leads to consistent results for the power counting
\cite{Barford:2004fz,Birse:2008wt,Ando:2008jb}.

Three-body calculations with external currents are still in their
infancy. However, a few exploratory calculations have been carried
out.  Universal properties of the triton charge form factor were
investigated in Ref.~\cite{Platter:2005sj} and neutron-deuteron
radiative capture was calculated in
Refs.~\cite{Sadeghi:2005aa,Sadeghi:2006aa,Sadeghi:2007qy}.
Electromagnetic properties of the triton were recently investigated in
Refs.~\cite{Sadeghi:2009dm,Sadeghi:2009rf}.  This work opens the
possibility to carry out accurate calculations of electroweak
reactions at very low energies for astrophysical processes.

The pionless approach has also been extended to the four-body sector
\cite{Platter:2004qn,Platter:2004zs}. In order to be able to apply the
Yakubovsky equations, an equivalent effective quantum mechanics
formulation was used. The study of the cutoff dependence of the
four-body binding energies revealed that no four-body parameter is
required for renormalization at leading order.  As a consequence,
there are universal correlations in the four-body sector which are
also driven by the large scattering length.  The best known example is
the Tjon line: a correlation between the triton and alpha particle
binding energies, $B_t$ and $B_\alpha$, respectively. Of course,
higher order corrections break the exact correlation and generate a
band.
\begin{figure}[tb!]
\centerline{\includegraphics*[width=8cm,angle=0]{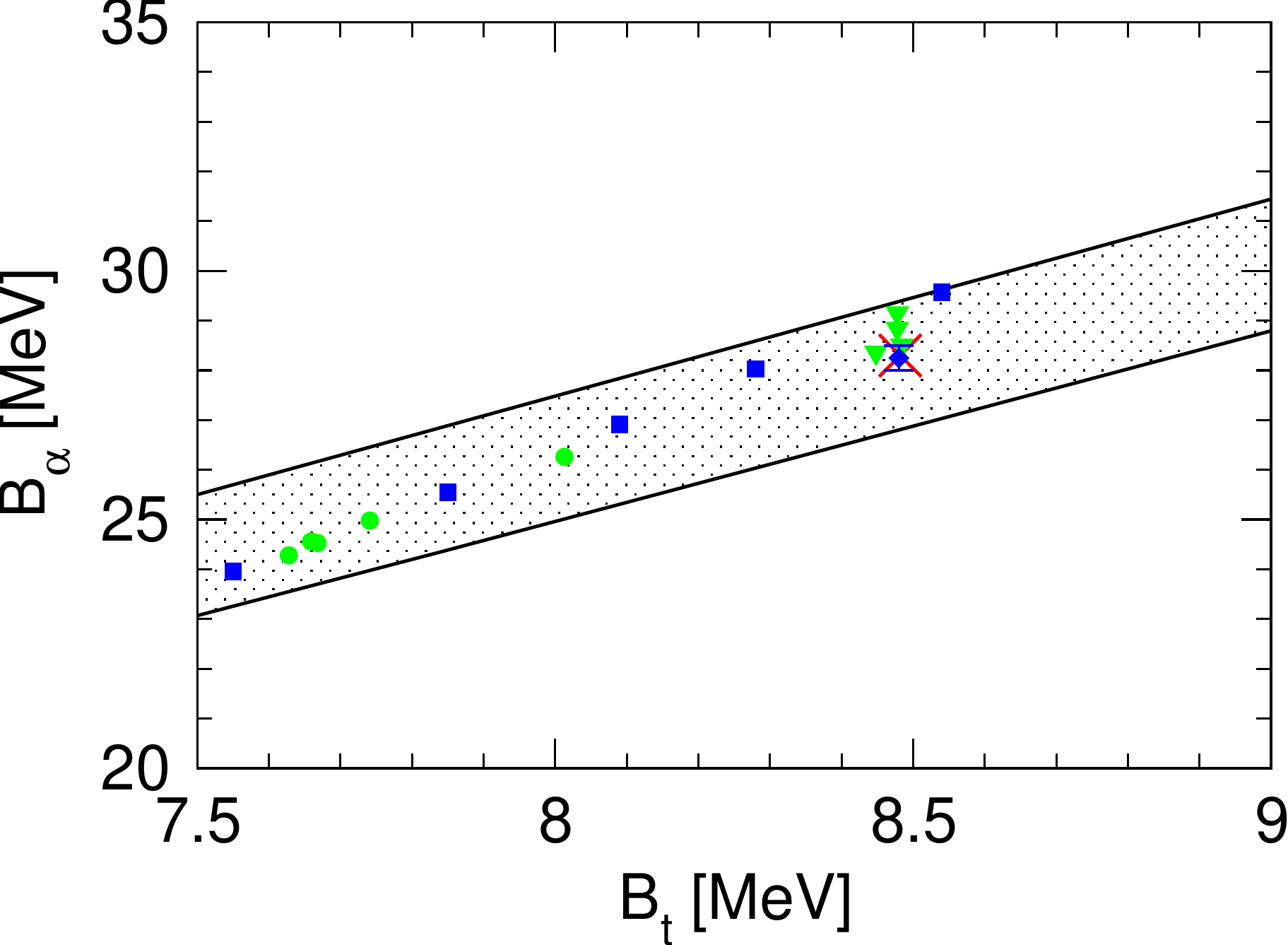}}
\caption{\label{fig:tjon}
\baselineskip 12pt 
The Tjon line correlation as predicted by the
pionless theory. The grey circles and triangles show
various calculations using phenomenological potentials \cite{Nogga:2000uu}.
The squares show the results of chiral EFT at NLO for different cutoffs
while the diamond gives the N$^2$LO result 
\cite{Epelbaum:2000mx,Epelbaum:2002vt}.
The cross shows the experimental point.
}
\end{figure}
In Fig.~\ref{fig:tjon}, we show this band together with some
calculations using phenomenological potentials \cite{Nogga:2000uu} and
a chiral EFT potential with explicit pions
\cite{Epelbaum:2000mx,Epelbaum:2002vt}.  All calculations with
interactions that give a large scattering length must lie within the
band. Different short-distance physics and/or cutoff dependence should
only move the results along the band. This can for example be observed
in the NLO results with the chiral potential indicated by the squares
in Fig.~\ref{fig:tjon} or in the few-body calculations with the
low-momentum NN potential $V_{\rm low\: k}$ carried out in
Ref.~\cite{Nogga:2004ab}. The $V_{\rm low\: k}$ potential is obtained
from phenomenological NN interactions by integrating out high-momentum
modes above a cutoff $\Lambda$ but leaving two-body observables (such
as the large scattering lengths) unchanged. The results of few-body
calculations with $V_{\rm low\: k}$ are not independent of $\Lambda$
but lie all close to the Tjon line (cf. Fig.~2 in
Ref.~\cite{Nogga:2004ab}).

Another interesting development is the application of the Resonating
Group Model to solve the pionless EFT for three- and four-nucleon
systems \cite{Kirscher:2009aj}. This method allows for a
straightforward inclusion of Coulomb effects. Kirscher et al.
extended previous calculations in the four-nucleon system to
next-to-leading order and showed that the Tjon line correlation
persists. Moreover, they calculated the correlation between the singlet
S-wave $^3$He-neutron scattering length and the triton binding
energy. Preliminary results for the halo nucleus $^6$He have been
reported in \cite{Kirscher:2009jw}.

The pionless theory has also been applied within the no-core shell
model approach. Here the expansion in a truncated harmonic oscillator
basis is used as the ultraviolet regulator of the EFT. The effective
interaction is determined directly in the model space, where an exact
diagonalization in a complete many-body basis is performed. In
Ref.~\cite{Stetcu:2006ey}, the $0^+$ excited state of $^4$He and the
$^6$Li ground state were calculated using the deuteron, triton, alpha
particle ground states as input. The first $0^+$ excited state in
$^4$He is calculated within 10\% of the experimental value, while the
$^6$Li ground state comes out at about 70\% of the experimental value
in agreement with the 30 \% error expected for the leading order
approximation. These results are promising and should be improved if
range corrections are included.  Finally, the spectrum of trapped
three- and four-fermion systems was calculated using the same method
\cite{Stetcu:2007ms}. In this case the harmonic potential is physical
and not simply used as an ultraviolet regulator.  For an update on
this work, see \cite{Stetcu:2009ic}.
\subsection{Quark Mass Dependence and Infrared Limit Cycle} 
\label{sec:quark-mass-depend}
In the following, we discuss the possibility of an exact infrared
limit cycle and the Efimov effect in a deformed version of QCD with
quark masses slightly larger than their physical values.  The quark
mass dependence of the chiral NN interaction was calculated to
next-to-leading order (NLO) in the chiral counting in
Refs.~\cite{Beane:2002xf,Epelbaum:2002gb}.  At this order, the quark
mass dependence is synonymous to the pion mass dependence because of
the Gell-Mann-Oakes-Renner relation:
$M_\pi^2 = -(m_u + m_d) \langle 0 | \bar{u} u | 0 \rangle/F_\pi^2\,,$
where $\langle 0 | \bar{u} u | 0 \rangle \approx (-290 \mbox{ MeV})^3$
is the quark condensate. In the following, we will therefore refer
to the pion mass dependence instead of the quark mass dependence
which is more convenient for our purpose.
The pion mass dependence of the nucleon-nucleon scattering lengths in the 
$\trip$--$^3{\rm D}_1$ and $\sing$ channels
as well as the deuteron binding energy were calculated
in Refs.~\cite{Beane:2001bc,Beane:2002xf,Epelbaum:2002gb}.

In principle, the pion mass dependence of the
chiral NN potential is determined uniquely. 
However, the extrapolation away from the 
physical pion mass generates errors. The dominating source are the
constants  $\bar C_{S,T}$ and $\bar D_{S,T}$ 
which give the corrections to the LO contact terms $\propto M_\pi^2$
and cannot be determined
independently from fits to data at the physical pion mass.
A smaller effect is due to the error in the LEC $\bar d_{16}$, which
governs the pion  mass dependence of $g_A$.
Both effects generate
increasing uncertainties as one extrapolates away from the physical point.

In the calculation of Ref.~\cite{Epelbaum:2002gb}, 
the size of the two constants $\bar D_{S}$ and $\bar D_{T}$ was
constrained from naturalness arguments, assuming 
$-3 \leq F_\pi^2 \Lambda_\chi^2 \bar D_{S,T} \leq 3\,,$
where $\Lambda_\chi \simeq$ 1 GeV is the chiral symmetry breaking scale.
These bounds are in agreement  with resonance saturation 
estimates \cite{Epelbaum:2001fm}.
The constant $\bar d_{16}$ was varied in the range 
$\bar d_{16} = -0.91 \ldots -1.76$ GeV$^{-2}$ \cite{Fettes:2000fd}.
These ranges were
used to estimate the extrapolation errors of two-nucleon observables
like the deuteron binding energy and the spin-singlet and 
spin-triplet scattering lengths \cite{Epelbaum:2002gb}. 
In the chiral limit the deuteron binding energy was found to be
of natural size, $B_d \sim F_\pi^2/m \simeq 10$~MeV. Note, however, that 
if larger uncertainties in the LECs $\bar D_{S}$ and $\bar D_{T}$ are
assumed one cannot make a definite statement about the binding
of the deuteron in the chiral limit \cite{Beane:2001bc,Beane:2002xf}. 
For pion masses above the physical value, however, all calculations show 
similar behavior.

\begin{figure}[tb!]
\centerline{
\includegraphics*[width=8cm,clip=true]{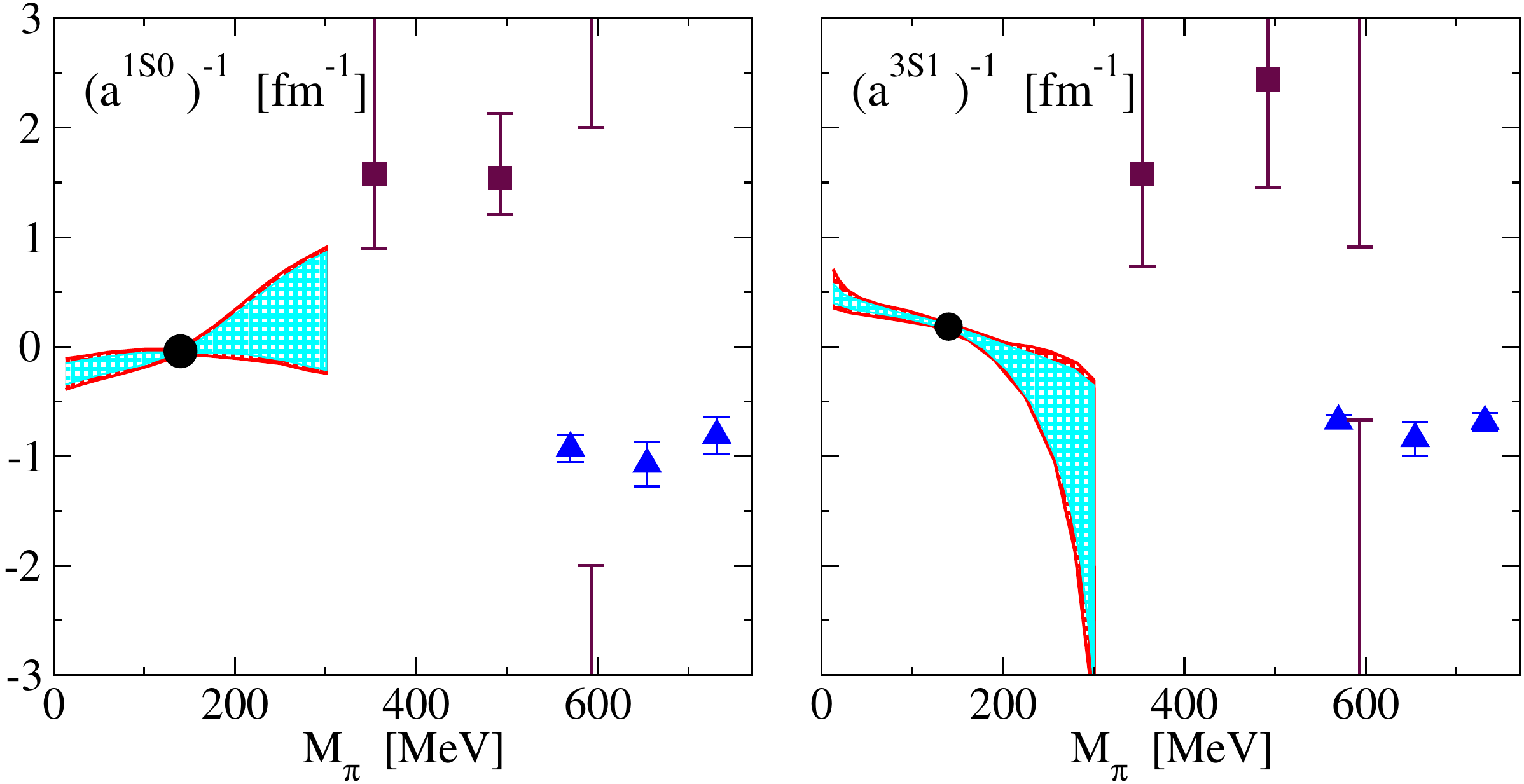}}  
\caption{
\baselineskip 12pt  
Inverse of the S-wave scattering lengths in the  
spin-triplet and spin-singlet nucleon-nucleon channels 
as a function of the pion mass $M_\pi$. Filled triangles and rectangles
show the lattice calculations from
Refs.~\cite{Fukugita:1994na,Fukugita:1994ve} and \cite{Beane:2006mx},
respectively.
}
\label{fig:asat}
\end{figure}
In Fig.~\ref{fig:asat}, we show
the inverse scattering lengths in the spin-triplet and  spin-singlet 
channels  from Ref.~\cite{Epelbaum:2002gb}
together with some recent lattice results \cite{Beane:2006mx}.
Figure \ref{fig:asat} also shows that a scenario where 
both inverse scattering lengths vanish simultaneously
at a critical pion mass of about $200$ MeV
is possible. For pion masses below the critical value, the spin-triplet
scattering length is positive and the deuteron is bound. 
As the inverse spin-triplet scattering length decreases, 
the deuteron becomes more and more
shallow and finally becomes unbound at the critical mass. 
Above the critical pion mass the deuteron exists as a shallow virtual 
state.  In the spin-singlet channel, the situation is reversed: the 
\lq\lq spin-singlet deuteron'' is a virtual state below the 
critical pion mass and becomes bound above. 
Based on this behavior Braaten ans Hammer conjectured that 
one should be able to reach the critical point by varying  
the up- and down-quark masses $m_u$ and $m_d$ independently
because the spin-triplet and spin-singlet channels have different isospin
\cite{Braaten:2003eu}.
In this case, the triton would display the Efimov effect which
corresponds to the occurrence of an infrared limit cycle in QCD.
It is evident that a complete investigation 
of this issue requires the inclusion
of isospin breaking corrections and therefore higher orders in the chiral
EFT. However, the universal 
properties of the limit cycle have been investigated 
by considering specific values of $\bar D_{S}$ and $\bar D_{T}$
that lie within the naturalness bound 
and cause the spin-singlet and 
spin-triplet scattering lengths to become infinite at the same value of 
the pion mass.

\begin{figure}[tb!]
\centerline{\includegraphics*[width=8cm,angle=0,clip=true]{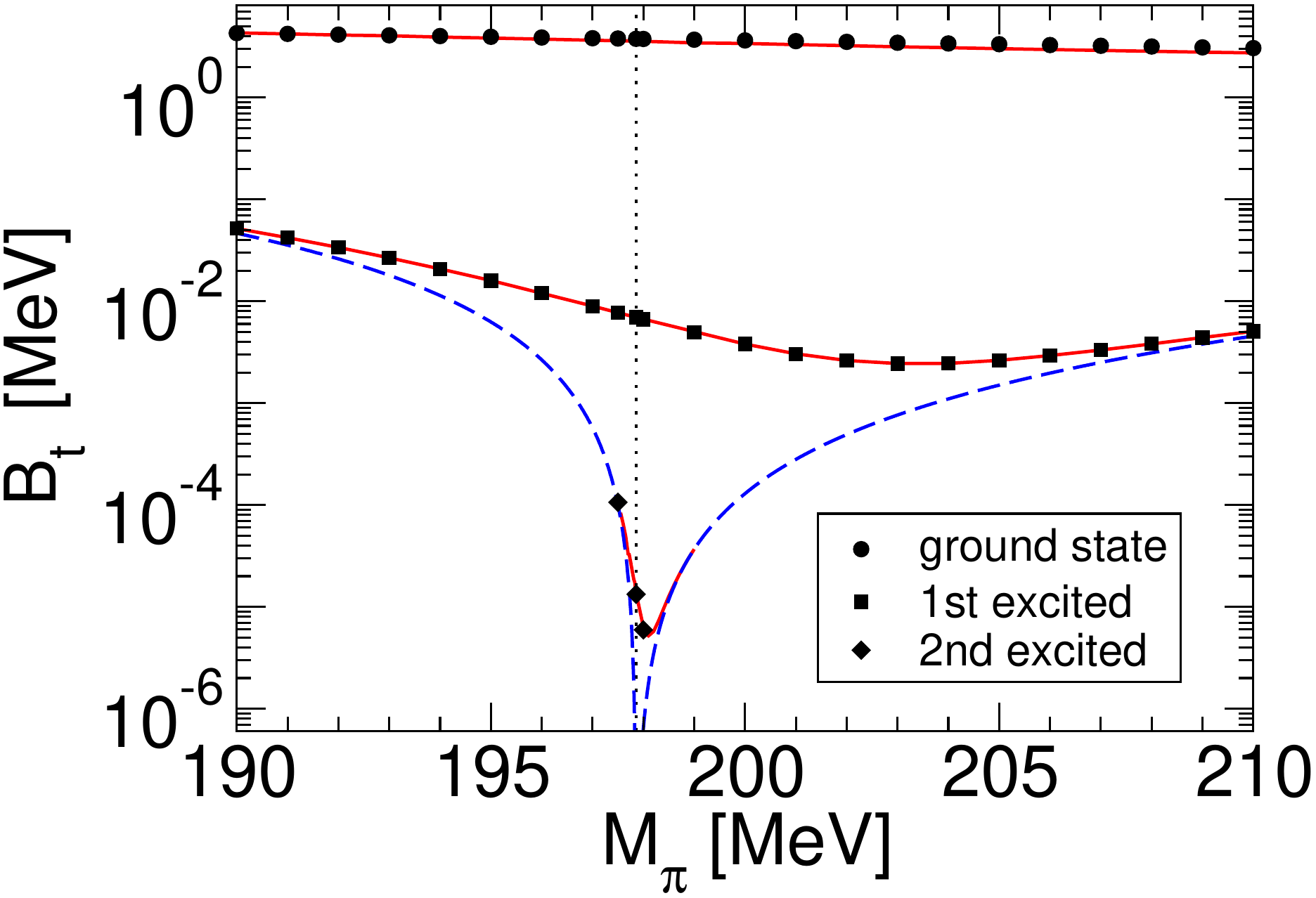}}
\caption{
\baselineskip 12pt  
Binding energies $B_t$ of the triton ground and first two
excited states as a function of $M_\pi$.
The circles, squares, and diamonds give the chiral EFT result, while
the solid lines are calculations in the pionless theory.
The vertical dotted line indicates the critical pion mass $\mpic$
and the dashed lines are the bound state thresholds.
}
\label{fig:bind3}
\end{figure}

In Ref.~\cite{Epelbaum:2006jc}, the properties of the triton around
the critical pion mass were studied for one particular solution with 
a critical pion mass $\mpic=197.8577$ MeV. 
{}From the solution of the Faddeev equations, 
the binding energies of the triton and the first
two excited states in the vicinity of the limit cycle
were calculated for this scenario in chiral EFT.
The binding energies are given in Fig.~\ref{fig:bind3}
by the circles (ground state), squares (first excited state),
and diamonds (second excited state). The dashed lines indicate the
neutron-deuteron ($M_\pi \leq \mpic$) and 
neutron-spin-singlet-deuteron ($M_\pi \geq \mpic$) thresholds
where the three-body states become unstable. Directly at 
the critical mass, these thresholds coincide with the three-body
threshold and the triton has infinitely many excited states.
The solid lines are leading order calculations in the pionless theory using 
the pion mass dependence of the nucleon-nucleon scattering lengths
and one triton state from chiral EFT as input.
The chiral EFT results for the other triton states in the critical region 
are reproduced very well.
The binding energy of the triton ground state
varies only weakly over the whole range of pion masses and is about 
one half of the physical value at the critical point. The excited states are
strongly influenced by the thresholds and vary much more strongly.

These studies were extended to N$^2$LO in the pionless EFT and 
neutron-deuteron scattering observables in Ref.~\cite{Hammer:2007kq}.
It was demonstrated that the higher order corrections in the vicinity
of the critical pion mass are small. 
\begin{figure}[tb!]
\centerline{\includegraphics*[width=8cm,angle=0,clip=true]{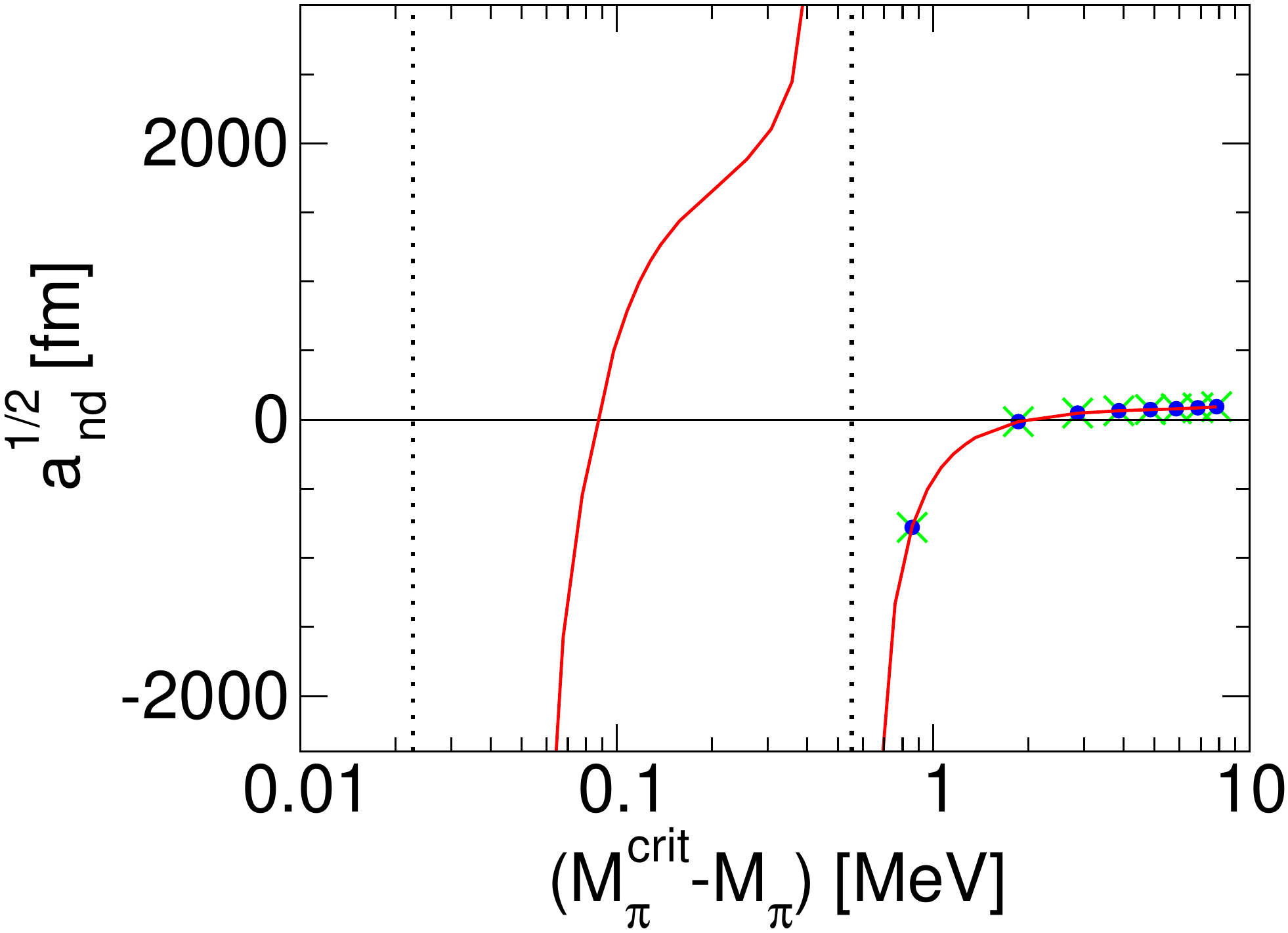}}
\caption{
\baselineskip 12pt  
Doublet neutron-deuteron scattering length $a_{nd}^{1/2}$
in the critical region computed in the pionless EFT. The solid line gives 
the LO result, while the crosses and circles show the NLO and 
N$^2$LO results. The dotted lines indicate the 
pion masses at which $a_{nd}^{1/2}$ diverges.
}
\label{fig:aND_crit}
\end{figure}
This is illustrated in Fig.~\ref{fig:aND_crit}, where
we show the doublet scattering length $a_{nd}^{1/2}$ in the critical
region.  The solid line gives the LO result, while the crosses and
circles show the NLO and N$^2$LO results. The dotted lines indicate
the pion masses at which $a_{nd}^{1/2}$ diverges because the second
and third excited states of the triton appear at the neutron-deuteron
threshold.  These singularities in $a_{nd}^{1/2}(M_\pi)$ are a clear
signature that the limit cycle is approached in the critical
region.  

A final answer on the question of whether an infrared limit cycle can
be realized in QCD can only be given by solving QCD directly. In particular,
it would be very interesting to know whether this can be achieved
by appropriately tuning the quark masses in a Lattice QCD simulation
\cite{Wilson:2004de}. The first full lattice
QCD calculation of nucleon-nucleon scattering was reported in
\cite{Beane:2006mx} but statistical noise presented a serious challenge.
A promising recent high-statistics study of three-baryon systems 
presented also initial results for a system with the quantum numbers 
of the triton such that lattice QCD calculations of three-nucleon 
systems are now within sight
\cite{Beane:2009gs}. For a review of these activities, see
Ref.~\cite{Beane:2008dv}.
Such calculations require a detailed understanding
of the modification of the Efimov spectrum in a cubic box. 
For identical bosons, there are significant finite volume shifts
even for moderate box sizes \cite{Kreuzer:2008bi}. These shifts show
universal scaling behavior which could be exploited to reduce the
computational effort \cite{Kreuzer:2009jp}. 
The extension of these studies to the triton case is in progress.

\subsection{Halo Nuclei}
\label{sec:halo-nuclei}
A special class of nuclear systems exhibiting universal behavior are
{\it halo nuclei} \cite{Zhukov-93, Jensen-04}.  Halo nuclei consist of
a tightly bound core surrounded by one or more loosely bound valence
nucleons. The valence nucleons are characterized by a very low
separation energy compared to those in the core.  As a consequence,
the radius of the halo nucleus is large compared to the radius of the
core. A trivial example is the deuteron, which can be considered a
two-body halo nucleus. The root mean square radius of the deuteron is
about three times larger than the size of the constituent nucleons.
Halo nuclei with two valence nucleons are particularly interesting
examples of three-body systems.  If none of the two-body subsystems are
bound, they are called {\it Borromean} halo nuclei.  This name is
derived from the heraldic symbol of the Borromeo family of Italy,
which consists of three rings interlocked in such way that if any one
of the rings is removed the other two separate.  The most carefully
studied Borromean halo nuclei are $^6$He and $^{11}$Li, which have two
weakly bound valence neutrons \cite{Zhukov-93}.  In the case of
$^6$He, the core is a $^4$He nucleus, which is also known as the
$\alpha$ particle.  The two-neutron separation energy for $^6$He is
about 1 MeV, small compared to the binding energy of the $\alpha$
particle which is about 28 MeV. The neutron-$\alpha$ ($n\alpha$)
system has no bound states and the $^6$He nucleus is therefore
Borromean. There is, however, a strong P-wave resonance in the
$J=3/2$ channel of $n \alpha$ scattering which is sometimes referred
to as $^5$He.  This resonance is responsible for the binding of
$^6$He. Thus $^6$He can be interpreted as a bound state of an
$\alpha$-particle and two neutrons, both of which are in $P_{3/2}$
configurations.

Because of the separation of scales in halo nuclei, they can be
described by extensions of the pionless EFT. One can assume the core
to be structureless and treats the nucleus as a few-body system of the
core and the valence nucleons.  Corrections from the structure of the
core appear in higher orders and can be included in perturbation
theory. Cluster models of halo nuclei then appear as leading order
approximations in this \lq\lq halo EFT''.  A new facet is the
appearance of resonances as in the neutron-alpha system which leads to
a more complicated singularity structure and renormalization compared
to the few-nucleon system discussed above \cite{Bertulani:2002sz}.

The first application of effective field theory methods to halo nuclei 
was carried out in Refs.~\cite{Bertulani:2002sz,Bedaque:2003wa}, where the 
$n\alpha$ system (``$^5$He'') was considered. It was found that 
for resonant P-wave interactions both the scattering length and effective 
range have to be resummed at leading order. At threshold, however, only 
one combination of coupling constants is fine-tuned and the EFT becomes
perturbative. Because the $n\alpha$ interaction is resonant in the P-wave 
and not in the S-wave, the binding mechanism of $^6$He is not the Efimov
effect. However, this nucleus can serve as a laboratory for studying
the interplay of resonance structures in higher partial waves. 

Three-body halo nuclei composed of a core and two valence neutrons are
of particular interest due to the possibility of these systems to
display the Efimov effect \cite{Efimov-70}. Since the scattering
length can not easily be varied in halo nuclei, one looks for Efimov
scaling between different states of the same nucleus. Such analyses
assume that the halo ground state is an Efimov state.\footnote{We note
  that it is also possible that only the excited state is an Efimov
  state while the ground state is more compact. This scenario can not
  be ruled out but is also less predictive.} They have previously been
carried out in cluster models and the renormalized zero-range model
(See, e.g. Refs.~\cite{Federov:1994cf,Amorim:1997mq,Mazumdar:2000dg}).
A comprehensive study of S-wave halo nuclei in halo EFT was recently
carried out in Ref.~\cite{Canham:2008jd}.  This work provided binding
energy and structure calculations for various halo nuclei including
error estimates.  Confirming earlier results by Fedorov et
al.\cite{Federov:1994cf} and Amorim et al.~\cite{Amorim:1997mq},
$^{20}$C was found to be the only candidate nucleus for an excited
Efimov state assuming the ground state is also an Efimov state.  This
nucleus consists of a $^{18}$C core with spin and parity quantum
numbers $J^P=0^+$ and two valence neutrons.  The nucleus $^{19}$C is
expected to have a $\frac{1}{2}^+$ state near threshold, implying a
shallow neutron-core bound state and therefore a large neutron-core
scattering length. The value of the $^{19}$C energy, however, is not
known well enough to make a definite statement about the appearance of
an excited state in $^{20}$C. An excited state with a binding energy
of about 65~keV is marginally consistent with the current experimental
information.

\begin{figure}[tb!]
\centerline{\includegraphics*[width=8cm,angle=0]{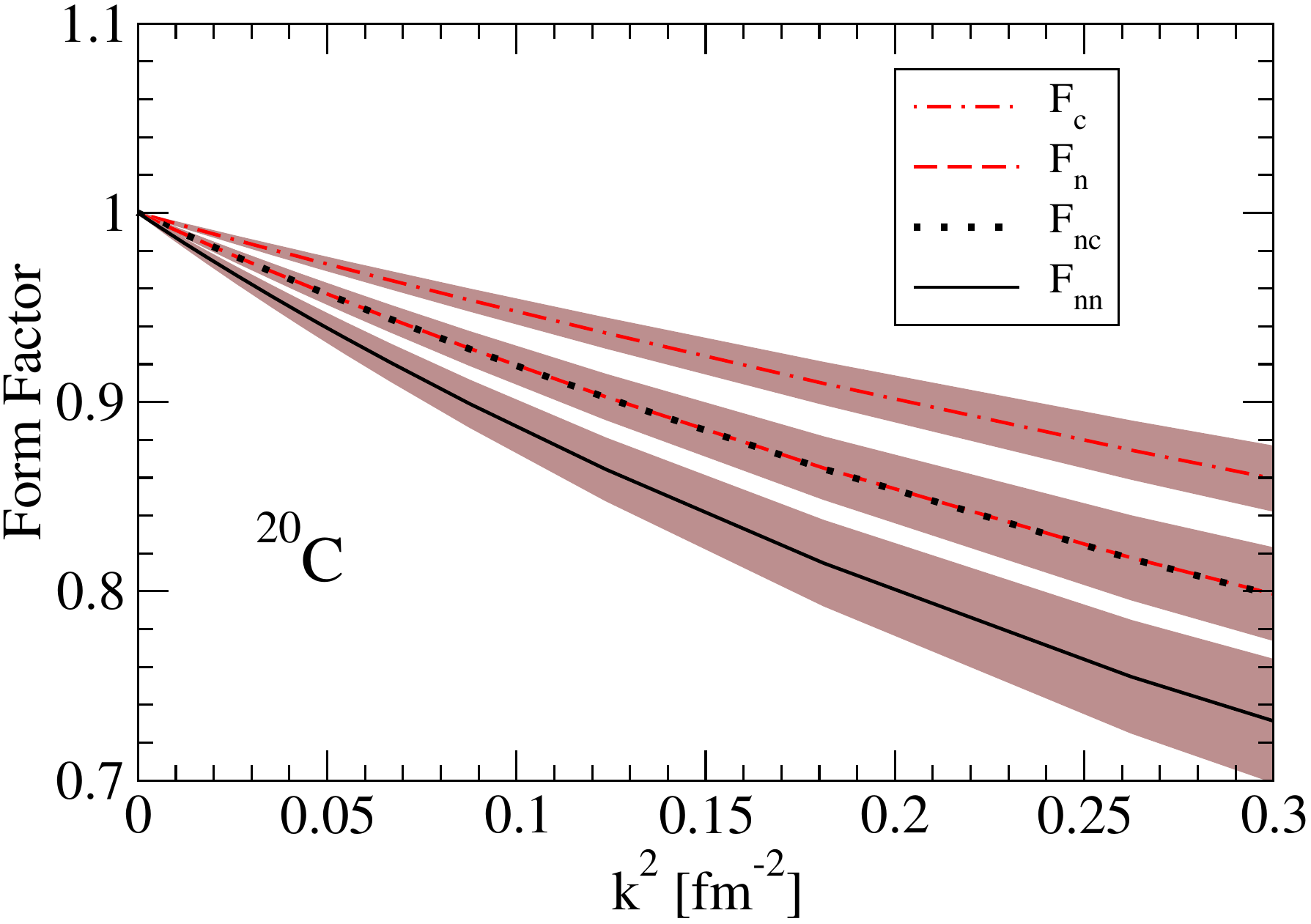}}
\caption{
\baselineskip 12pt  
The one- and two-body matter density form factors 
${\cal F}_{c}$, ${\cal F}_{n}$, ${\cal F}_{nc}$, and 
${\cal F}_{nn}$ with leading order error bands for the ground 
state of $^{20}$C as a function of the momentum transfer $k^2$.}
\label{fig:spec.halo}
\end{figure}

The matter form factors and radii of halo nuclei can also be
calculated in the halo EFT \cite{Yamashita:2004pv,Canham:2008jd}.  As
an example, we show the various one- and two-body matter density form
factors ${\cal F}_{c}$, ${\cal F}_{n}$, ${\cal F}_{nc}$, and ${\cal
  F}_{nn}$
with leading order error bands for the ground state of $^{20}$C as a function
the momentum transfer $k^2$ from \cite{Canham:2008jd}
in Fig.~\ref{fig:spec.halo}.
The theory breaks down for momentum transfers of the order of the pion-mass 
squared ($k^2\approx 0.5$ fm$^{-2}$). 

{}From the slope of the matter form factors one can extract the 
corresponding radii:
\beq
{\mathcal F}(k^2) =  1 
- {1 \over 6} k^2 \left\langle r^2 \right\rangle + \ldots\, .
\label{FFexpand}
\eeq Information on these radii has been extracted from experiment for
some halo nuclei based on intensity interferometry and Dalitz plots
\cite{Marques:2001pe}. Within the error estimates, the extracted
values are in good agreement with the theoretical predictions from
halo EFT \cite{Canham:2008jd}.  For the possible $^{20}$C excited
state, the halo EFT at leading order predicts neutron-neutron and
neutron-core radii of order 40~fm while the ground state radii are of
order 2-3~fm.  The theoretical errors are estimated to be of order
10\%.  Assuming a natural value for the effective range of the
$n$-$^{18}$C interaction, $r_0\approx 1/M_\pi$, next-to-leading order
predictions for these radii have recently been obtained
\cite{Canham:2009xg}. The leading order results were found to be
stable under inclusion of the leading effective range corrections and
the typical errors could be reduced to about 1-2\%.

Scattering observables offer a complementary window on Efimov physics
in halo nuclei and some recent model studies have focused on this
issue.  In particular, in
Refs.~\cite{Yamashita:2007ej,Mazumdar:2006tn} the trajectory of the
possible $^{20}$C excited state was extended into the scattering
region in order to find a resonance in $n$-$^{19}$C scattering.  A
detailed study of $n$-$^{19}$C scattering near an Efimov state was
carried out in \cite{Yamashita:2008sg}.

\begin{figure}[t!]
\centerline{\includegraphics*[width=8cm,angle=0,clip=true]{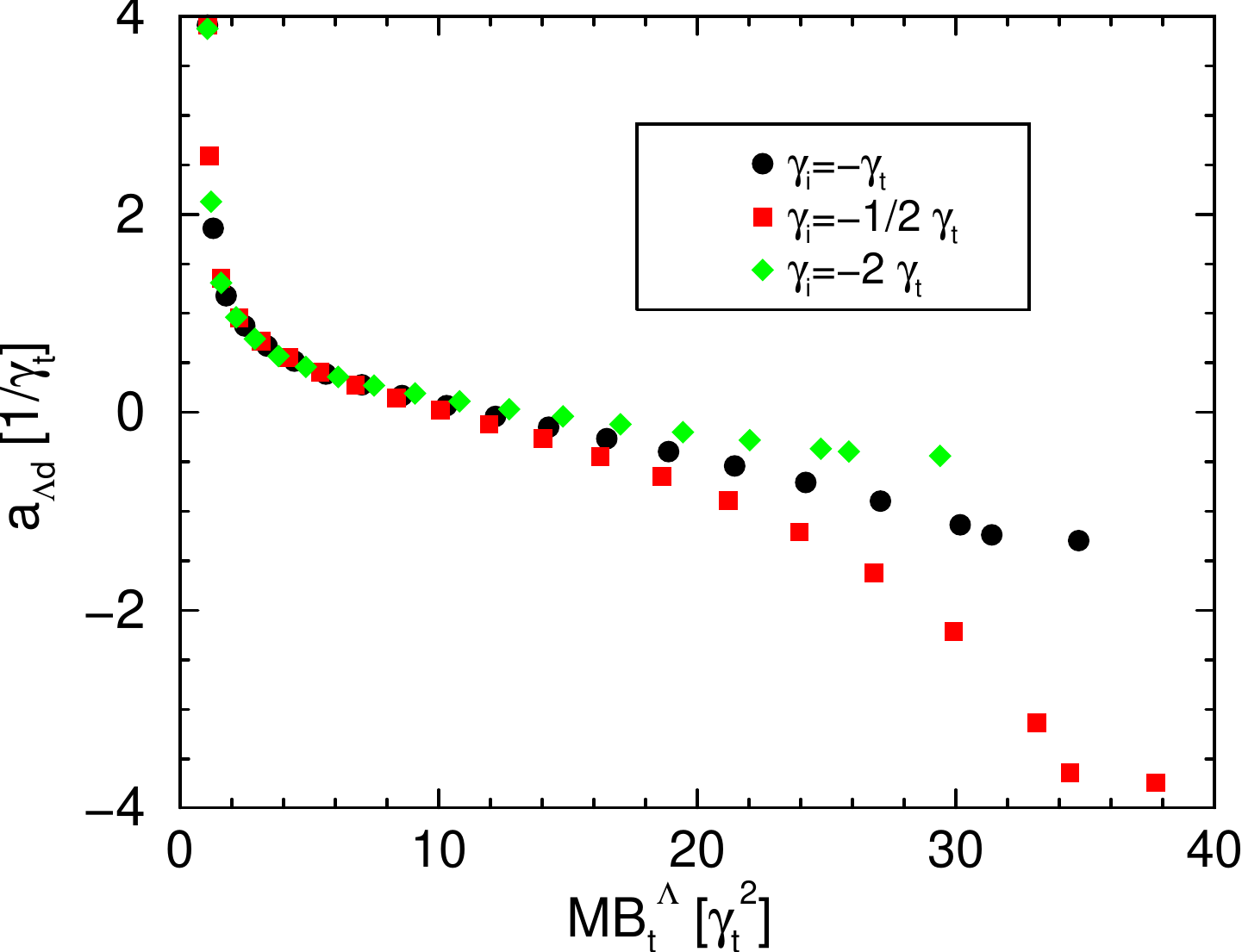}}
\caption{
\baselineskip 12pt  
Phillips line in the hypertriton channel 
for different values of the $\Lambda N$ pole position
$\gamma_i$ (all quantities are in units of the deuteron pole position
$\gamma_t\approx 45$ MeV).}
\label{fig:phil}
\end{figure}

The simplest strange halo nucleus is the hypertriton, a three-body
bound state of a proton, neutron and  the $\Lambda$. 
The total binding energy is only about 2.4 MeV. 
The separation energy for the $\Lambda$, $E_\Lambda = 0.13$ MeV, 
is tiny compared to the binding energy $B_d = 2.22$ MeV of the 
deuteron.  The hypertriton can therefore also be considered a two-body 
halo nucleus.
It has been studied in both two-body and three-body approaches
\cite{Cobis:1996ru,Fedorov:2001wj,Congl92}.
A study of the hypertriton in the halo EFT was carried out in
Ref.~\cite{Hammer:2001ng}.
The $\Lambda N$ scattering lengths are not well known experimentally
since the few scattering data are at relatively high energies.
If the $\Lambda N$ scattering lengths are large,
the hypertriton is likely bound due to the Efimov effect.
In this case there will also be a correlation between the $\Lambda d$ 
scattering length $a_{\Lambda d}$ and the hypertriton binding energy 
$B_t^\Lambda$ analog to the Phillips line in the neutron-deuteron
system \cite{Fedorov:2001wj}. 
In Fig.~\ref{fig:phil}, we show this Phillips line correlation 
for three values of  the $\Lambda N$ pole position $\gamma_i$
\cite{Hammer:2001ng}. 
For small hypertriton energies $B_t^\Lambda$, the different Phillips lines
coincide exactly (the physical hypertriton corresponds
to $MB_t^\Lambda \approx 1.06\,\gamma_t^2$) and deviate from each other 
only at very large binding energies where the EFT breaks down. 
For all practical purposes the Phillips line is therefore independent
of $\gamma_i$. 
Whether the Efimov effect plays a role for the hypertriton is an open
question.
Most modern hyperon-nucleon potentials, however, favor a natural
$\Lambda N$ scattering length \cite{Epelbaum:2008ga}. 

Another powerful method that can be used to investigate the Efimov
effect in three-body halo nuclei at existing and future facilities
with exotic beams (such as FAIR and FRIB) is Coulomb excitation.  In
these experiments a nuclear beam scatters off the Coulomb field of a
heavy nucleus. Such processes can populate excited states of the
projectile which subsequently decay, leading to its ``Coulomb
dissociation'' \cite{Bert88}. The halo EFT offers a systematic
framework for a full quantum-mechanical treatment and can be used to
predict the signature of the Efimov effect in these reactions.

\subsection{Three-Alpha System and Coulomb Interaction}
The excited $0^+$ state in $^{12}$C is known as the Hoyle state.
Its properties are important for stellar astrophysics since it
determines the ratio of carbon to oxygen in stellar helium burning.
Efimov suggested 
that the Hoyle state
could be explained as an Efimov state of $\alpha$ particles
\cite{Efimov-70,Efimov:1971zz}.
(For a more detailed discussion, see Ref.~\cite{Efimov90}.)
In this case, the universal properties are modified by 
the long-range Coulomb interaction. The
modified Efimov spectrum needs to be understood before any definite
statement about the nature of the Hoyle state can be made.

Many recent studies have focused on the consistent inclusion of the
Coulomb interaction in two-body halo nuclei such as the $p\alpha$ and
$\alpha\alpha$ systems \cite{Higa:2008rx,Higa:2008dn}.  In particular,
the $\alpha\alpha$ system shows a surprising amount of fine-tuning
between the strong and electromagnetic interaction. It can be
understood in an expansion around the limit where, when
electromagnetic interactions are turned off, the $^8$Be ground state
is exactly at threshold and exhibits conformal invariance
\cite{Higa:2008dn}. In this
scenario, the Hoyle state in $^{12}$C would indeed appear as a remnant
of an excited Efimov state.  In order to better
understand the modification of the Efimov spectrum and limit cycles by
long-range interactions such as the Coulomb interaction, a one
dimensional inverse square potential supplemented with a Coulomb
interaction was investigated in \cite{Hammer:2008ra}.  The results
indicate that the counterterm required to renormalize the inverse
square potential alone is sufficient to renormalize the full
problem. However, the breaking of the discrete scale invariance
through the Coulomb interaction leads to a modified bound state
spectrum. The shallow bound states are strongly influenced by the
Coulomb interaction while the deep bound states are dominated by the
inverse square potential.  These results support the conjecture of the
Hoyle state being an Efimov state of $\alpha$ particles but a full
calculation of the $3\alpha$ system including Coulomb in the halo EFT
is missing. Calculations with the fermionic molecular dynamics
model and electron scattering data, however, support a 
pronounced $\alpha$ cluster structure of the Hoyle state 
\cite{Chernykh:2007zz}.

\section{Applications in Particle Physics}
\subsection{Hadronic Molecules}
In recent years many new and possibly exotic charmonium 
states have been observed at the 
B-factories at SLAC, at KEK in Japan, and at the CESR collider at Cornell.
This has revived the field of charmonium spectroscopy
\cite{Eichten:2007qx,Voloshin:2007dx,Godfrey:2008nc}.
Several of the new states exist very close to scattering thresholds,
and can be interpreted as hadronic molecules.
If they are sufficiently shallow, one may ask whether there are any 
three-body hadronic molecules bound by the Efimov effect. 

A particularly interesting example is the $X(3872)$, discovered by the
Belle collaboration \cite{Choi:2003ue} in $B^{\pm}\to K^\pm \pi^+
\pi^- J/\psi$ decays and quickly confirmed by CDF
\cite{Acosta:2003zx}, D0 \cite{Abazov:2004kp}, and BaBar
\cite{Aubert:2004ns}.  The state has likely quantum numbers
$J^{PC}=1^{++}$ and is very close to the $D^{*0} \bar{D}^0$ threshold.
As a consequence, the $X(3872)$ has a resonant 
S-wave coupling to the $D^{*0} \bar{D}^0$ system.
An extensive program
provides predictions for its decay modes based on the assumption that
it is a  $D^{*0} \bar{D^0}$ molecule with even C-parity:
\beq
(D^{*0} \bar{D}^0)_+ \equiv \frac{1}{\sqrt{2}}
\left(D^{*0} \bar{D}^0+D^{0} \bar{D}^{*0}\right)\,.
\label{eq:Xflavor}
\eeq
This assumption naturally explains several puzzling features
such the apparently different mass in the 
$J/\psi \pi^+ \pi^-$ and $D^{*0} \bar{D}^0$ decay channels and
the isospin violating decays \cite{Braaten:2007dw,Braaten:2007ft}. 
A status report with references
to the original literature can be found in 
Ref.~\cite{Braaten:2008nv}.

Using the latest measurements in the  $J/\psi \pi^+ \pi^-$ channel, 
the mass of the $X(3872)$ is \cite{Canham:2009zq}:
$m_X = (3871.55 \pm 0.20) \mbox{ MeV}\,,$
which corresponds to an energy relative to the $D^{*0}\bar{D}^0$ threshold of
\beq
E_X =(-0.26 \pm 0.41) \mbox{ MeV}\,.
\label{eq:EX}
\eeq The central value corresponds to a $(D^{*0} \bar{D}^0)_+$ bound
state with binding energy $B_X=0.26$ MeV (but a virtual state cannot
be excluded from the current data in the $J/\psi \pi^+ \pi^-$ and
$D^{*0} \bar{D}^0$ channels
\cite{Bugg:2004rk,Hanhart:2007yq,Voloshin:2007hh}).  The $X(3872)$ is
also very narrow, with a width smaller than 2.3 MeV.

Because the $X(3872)$ is so close to the $D^{*0} \bar{D}^0$ threshold,
it has universal low-energy properties that depend only on its binding
energy \cite{Braaten:2003he}.  Close to threshold, the coupling to
charged $D$ mesons can be neglected because the $D^{*+} \bar{D}^-$
threshold is about 8 MeV higher in energy.  Therefore, the properties
of the $X(3872)$ can be described in a universal EFT with contact
interactions only.  Unfortunately, there is no Efimov effect in this
system \cite{Braaten:2003he}, so universal bound states of the $X$ and
$D^0$ or $D^{*0}$ mesons do not exist.  The reason for this that there
is not a sufficient number of pairs with resonant interactions as only
the $D^{*0} \bar{D}^0$ and $D^{0} \bar{D}^{*0}$ interactions are
resonant. However, it is possible to provide model-independent
predictions for the scattering of $D^0$ and $D^{*0}$ mesons and their
antiparticles off the $X(3872)$. This scattering process is to leading
order determined by the $D^{*0} \bar{D}^0$ and $D^{0} \bar{D}^{*0}$
interactions only.

\begin{figure}[t!]
	\centerline{\includegraphics*[width=9cm]{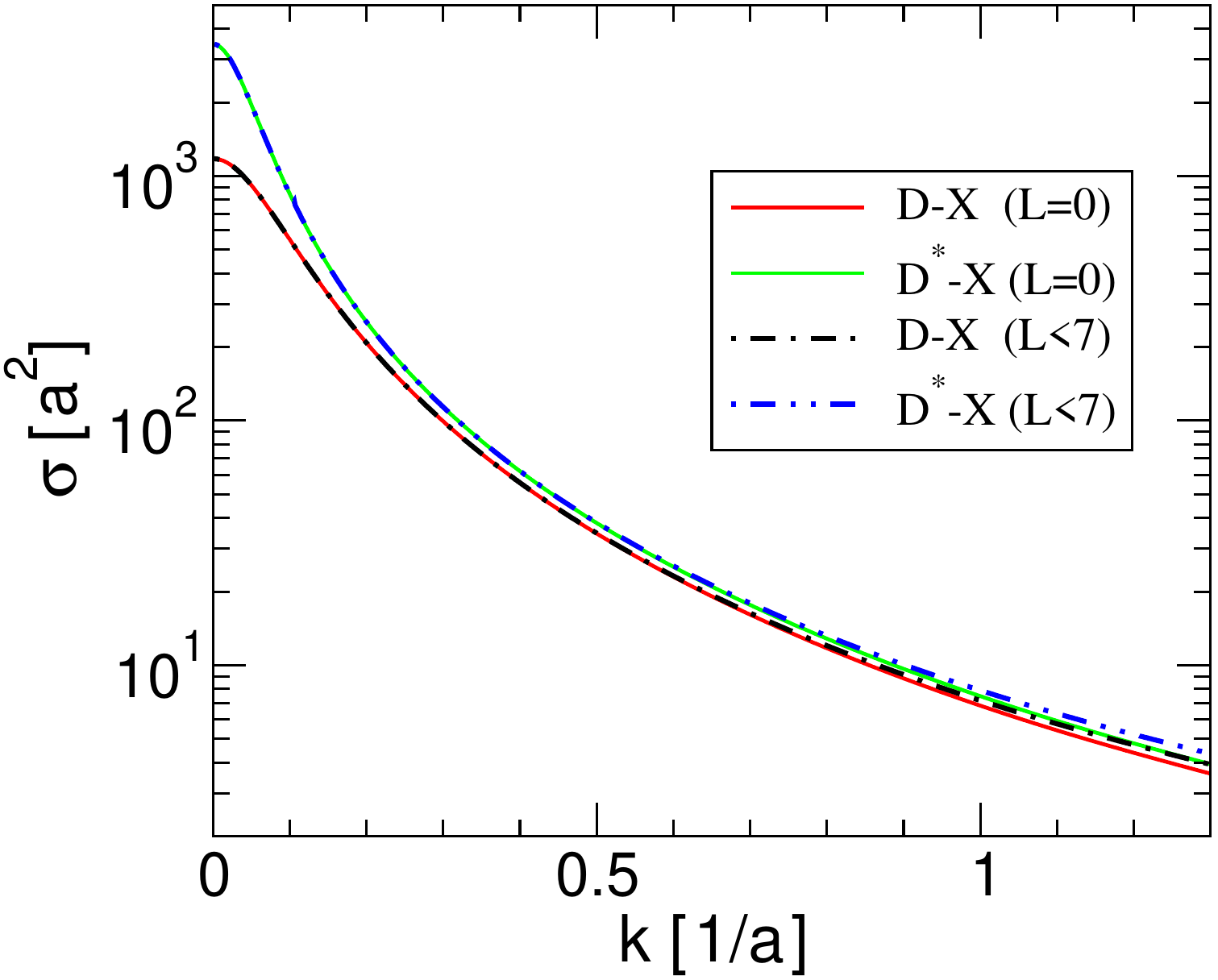}}
	\caption{
\baselineskip 12pt  
Total cross section for scattering of $D^0$ and $D^{*0}$ 
          mesons off the $X(3872)$ for S-waves ($L=0$) 
          and including higher partial waves with $L<7$,
          in units of the scattering length $a$.  
          The cross section is the same
          for the scattering of particles as it is for the scattering of 
antiparticles.}
	\label{fig:xsec}
\end{figure}

The corresponding cross sections as a function of the center-of-mass 
momentum $k$ obtained in \cite{Canham:2009zq} are shown in Fig.~\ref{fig:xsec}. 
The difference between the 
contribution of S-waves ($L=0$) and the full cross section
(including all partial waves up to $L=6$) is negligible 
for momenta below the bound state pole momentum $\gamma$. 
Our results are given 
in units of the scattering length and may be scaled
to physical units once $a$ is determined. At present 
the error in the experimental value for $E_X$ in Eq.~(\ref{eq:EX}) implies 
a large error in the scattering length. In particular, 
we obtain the ranges $\gamma=(0 \ldots 36)$ MeV 
for the pole momentum and 
$a=(5.5 \ldots \infty)$ fm for the scattering length
with central values  $\gamma=22$ MeV 
and $a=8.8$ fm. 
Using the central value of the scattering length,
we obtain for the scale factor $a^2= 0.78$ barn. This factor can
become infinite if the $X(3872)$ is directly at threshold, while the
lower bound from the error in $E_X$ would give a value of 0.3
barn. Even in this case the total cross section at threshold will be
of the order 300 barns for $D^0 X$ scattering and 1000 barns for
$D^{0*} X$ scattering.  It may be possible to extract the scattering
within the final state interactions of $B_c$ decays and/or other LHC
events. Observation of enhanced final state interactions would
provide an independent confirmation of the nature of the $X(3872)$.

There may also be hadronic three-body molecules that are bound due to the 
Efimov effect but currently no strong candidate states are known.
This situation will be improved by new experiments at facilities
such as FAIR and Belle II which have a dedicated program to 
study exotic charmonium states.

\section{Summary and Outlook}
\label{sec:summary}
Any few-body system with short-range interactions 
that has a two-body scattering length larger than the
range of the underlying interaction will display universal
properties and Efimov physics.
This statement is independent of the typical length
scale of the system and atomic, nuclear and particle physics can
provide examples of universality. 
While much progress has been made recently in experiments with
ultracold atoms, the concept of Efimov physics was originally 
devised for the few-nucleon problem. As we have shown, 
it can serve as a starting
point for a description of very low-energy nuclear phenomena
in an expansion around the unitary limit. 

In this review we have discussed the manifestation of Efimov
physics with a strong emphasis on nuclear and particle physics.
There are a number of such systems that
display low-energy universality associated with Efimov physics.
The most important example
from nuclear physics is the triton. The nucleon-nucleon scattering
length is large compared to the range of the internuclear
interaction. Phase shift equivalent $NN$ potentials will therefore
necessarily give results for triton binding energy and
neutron-deuteron scattering length that are correlated and
lie on the Phillips line.
The implications of universality on the four-nucleon system have been
also been explored. It was found that the Tjon line, a correlation
between three-nucleon and four-nucleon binding energies is a result of
the large two-nucleon scattering length.

Halo nuclei might provide a further example of few-body
universality. In particular two-neutron Halos such as $^{20}$C could
display Efimov physics in form of an excited three-body state due to
the large core-neutron scattering length. While first studies of bound
state observables have become available, scattering
calculations represent an exciting opportunity for future
applications of the EFT approach. In particular so-called $p-t$
reactions in which a triton is formed in a collision of a proton and a
two-neutron halo will provide an important benchmark.

Several new charmonium states have recently been discovered close to
scattering thresholds and can be interpreted as hadronic molecules.  If
they are sufficiently shallow, they have universal properties
associated with large scattering length physics. The best known
example is the $X(3872)$ which may be interpreted as a $D^{*0}
\bar{D^0}$ molecule with even C-parity. There may also be three-body
hadronic molecules bound by the Efimov effect but currently no strong
candidates are known.

The separation of scales between scattering length and range
facilitates the application of an EFT that reproduces at leading order
the results obtained by Efimov and is known as the pionless
EFT. Within this framework corrections to the zero-range limit can be
calculated systematically in a small parameter expansion in powers of
$k \ell$ and $\ell/a$. The number of few-body calculations that
include higher order corrections is growing. As a consequence, the
expansion around the unitary limit provides a useful starting
point for a controlled description of very low energy phenomena
in nuclear and particle physics. The intricate consequences
of Efimov physics are explicit in this framework and universal
correlations between observables arise naturally.
Moreover, this theory is an ideal tool to unravel universal
properties and establish connections between different 
fields of physics.

The constituents of nuclear few-body systems can have charge in
contrast to the neutral atoms used in experiments with ultracold
gases. Electroweak observables provide thus additional information on
few-body universality. However, they are also of interest by
themselves and the pionless EFT guarantees a consistent framework
for the calculation of observables with the minimal number of
parameters as it is straightforward to consider external currents in
any EFT. First calculations for observables such as form factors and
capture rates have been performed but many more remain to be
calculated.  Of very high interest are in this context thermal capture
rates in the four-nucleon sector that are relevant to big bang
nucleosynthesis.

\section*{Acknowledgements}

We thank Eric Braaten and Vitaly Efimov for discussions.
This research was supported by the
National Science Foundation under Grant No.~PHY--0653312,
by the UNEDF SciDAC Collaboration under DOE Grant DE-FC02-07ER41457, 
by the Department of Energy under grant number DE-FG02-00ER41132,
by the Deutsche Forschungsgemeinschaft through SFB/TR16, and 
by the Bundesministerium f\"ur Bildung und Forschung
under contracts 06BN411 and 06BN9006.

\bibliographystyle{arnuke_revised}
\bibliography{/Users/lplatter/projects/References}
\end{document}